\documentclass[a4paper,fleqn]{cas-dc}

\usepackage[numbers]{natbib}
\usepackage{graphicx}
\usepackage{textcomp}
\usepackage{xcolor}
\usepackage{longtable}
\usepackage{lscape}
\usepackage{lipsum}
\usepackage[export]{adjustbox}

\def\tsc#1{\csdef{#1}{\textsc{\lowercase{#1}}\xspace}}
\tsc{WGM}
\tsc{QE}
\tsc{EP}
\tsc{PMS}
\tsc{BEC}
\tsc{DE}
\DeclareUnicodeCharacter{2212}{-}
\begin{document}
\rule{\textwidth}{1pt}
\let\WriteBookmarks\relax
\def\floatpagepagefraction{1}
\def\textpagefraction{.001}
\shorttitle{convergence of fog and vehicularsocial networks survey}
\shortauthors{Farimasadat Miri et al.}

\title [mode = title]{A Comprehensive Survey on the Convergence of Vehicular Social Networks and Fog Computing}                      



\author[]{Farimasadat Miri}

\fnmark[]
\ead{farimasadat.miri@uoit.ca}

\address[]{Ontario Tech University}

\author[]{Richard Pazzi}[]

\fnmark[]
\ead{richard.pazzi@ontariotechu.ca}



\begin{abstract}
In recent years, the number of IoT devices has been growing fast which leads to a challenging task for managing, storing, analyzing, and making decisions about raw data from different IoT devices, especially for delay-sensitive applications. In a vehicular network (VANET) environment, the dynamic nature of vehicles makes the current open research issues even more challenging due to the frequent topology changes that can lead to disconnections between vehicles. To this end, a number of research works have been proposed in the context of cloud and fog computing over the 5G infrastructure. On the other hand, there are a variety of research proposals that aim to extend the connection time between vehicles. Vehicular Social Networks (VSNs) have been defined to decrease the burden of connection time between the vehicles. This survey paper first provides the necessary background information and definitions about fog, cloud and related paradigms such as 5G and SDN. Then, it introduces the reader to Vehicular Social Networks, the different metrics and the main differences between VSNs and Online Social Networks. Finally, this survey investigates the related works in the context of VANETs that have demonstrated different architectures to address the different issues in fog computing. Moreover, it provides a categorization of the different approaches and discusses the required metrics in the context of fog and cloud and compares them to Vehicular social networks. A comparison of the relevant related works is discussed along with new research challenges and trends in the domain of VSNs and fog computing.
\end{abstract}

\begin{graphicalabstract}
\includegraphics{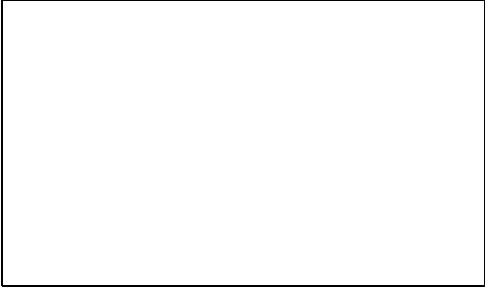}
\end{graphicalabstract}



\maketitle
\section{Introduction}
\label{sec:intro}
Over the past several decades, the number of computers and electronic devices has increased dramatically, leading to a commensurate need to connect these devices. Partially due to this dramatic increase, the proliferation of IoT devices has become unavoidable. Such IoT devices are able to generate massive amounts of data that require storage, management, processing, and production of metadata in a structured way. This massive generation of data has contributed to both the increase in attention given to big data analysis in recent years and the acceleration of IoT devices during the development of smart cities. In addition to IoT devices, vehicles also play an important role in generating and sensing data, and the volume of data generated from onboard sensors is likewise massive. However, unlike stationary IoT devices, the mobile nature of vehicles allows them to generate and sense data across large areas of cities. Like IoT devices, the data coming from onboard sensors is massive. In some cases, the vehicles need to be connected to each other in order to send important data. However, vehicle mobility prevents them from having seamless connectivity with each other. There are different approaches that tackle this issue in order to extend the vehicle connectivity. A number of organizations and companies are using cloud services to deal with big data analytics. Cloud environments come in different types, such as private and community clouds and have different services such as Software as a Service (SaaS), Infrastructure as a Service (IaaS) and Platform as a Service (PaaS). Although cloud provides powerful servers for storing and analyzing data, processing massive amounts of data from onboard sensors and other IoT devices in a timely manner, especially for real time applications, takes time and cannot satisfy user requirements in delay sensitive applications. A number of techniques has been proposed to solve this issue, and Fog computing is one of them. Thus, in order to suppress the deficiencies that currently exist in cloud environments, fog computing basically is set up near the IoT devices to cover the necessary needs of time critical applications. Through this architecture, we have computation, storage and decision-making tasks occurring not only in the clouds, but also close to the end devices as well. It’s important to point out that although fog can cover some drawbacks in cloud, it brings some new challenges with itself, such as synchronization between fog nodes and the lack of a standard to deal with heterogeneous environments. There are some related paradigms regarding fog computing, such as Edge computing, Mobile edge computing and so on, which will be discussed in this paper shortly.
\newline

Fog computing has many benefits, such as using idle resources for storing and processing data near edge devices. Not only can it meet delay-sensitive application requirements, but it can also help us decrease the cost of resource usage. While we are using fog computing for having better efficiency in our network, in the context of VANETs, we need to extend the duration of connections between cars as long as we can. To this end, Vehicular social network is one of the emerging topics in recent years that can help with connectivity problems in VANETs.

Vehicular Social network inspires fundamental ideas of online social network. However, Vehicular social network and online social network are not the same, and they have differences in some features which will be addressed in this paper.
A number of works in the literature have focused on fog definitions, requirements and related challenges. This paper goes beyond that and deeply investigates the recent related works in fog and vehicular social networks in the scope of 5G for VANETs. Then, we prepare some tables and compare them with one another to understand which of them have satisfied different terms in fog and VSN. To this end, this paper sheds light on the convergence of Fog and VSN technologies in the context of VANETs. It also investigates the integration with Social Defined Networks (SDNs) and the adoption of 5G infrastructures for VANETs.  we try to combine the idea of fog and vehicular social network that in many cases for having better control and data management, SDN (Software Defined Network) is part of it, coupled with 5G infrastructures.
The contribution of this paper is summarized as follows: 
\begin{itemize}
    \item A thorough investigation of the existing literature on Fog computing and its intersection with Social Networks for improving Vehicular Networks
    
    \item A comprehensive comparative study of the related works in fog and VSNs.
    \item Comparative study and analysis of the related works in Fog and VSN in the context of VANET.
    \item Discussion of the Recent challenges, issues and new trends in fog and Vehicular Social Network.
  
\end{itemize}



The remainder of this paper is organized as follows. Section \ref{sec:intro} provides a quick overview of the current challenges in IoT, Cloud, and fog. Fog definition, architecture in the fog for VANET, and Differences between cloud and fog have been covered in section \ref{sec:fog}. Preliminary terms in 5G and SDN have been demonstrated in section \ref{pre-term}. Section \ref{VSN} discusses Vehicular social networks, related metrics, and differences between Vehicular Social Network (VSN) and Online Social Network (OSN). Related works have been covered in section \ref{rel-work}. Research challenges and conclusions have been covered in section \ref{challenges} and \ref{conc}.

\section{Fog definition}
\label{sec:fog}
\begin{figure*}[t!]
\centering
\includegraphics[width=13cm, height=10cm]{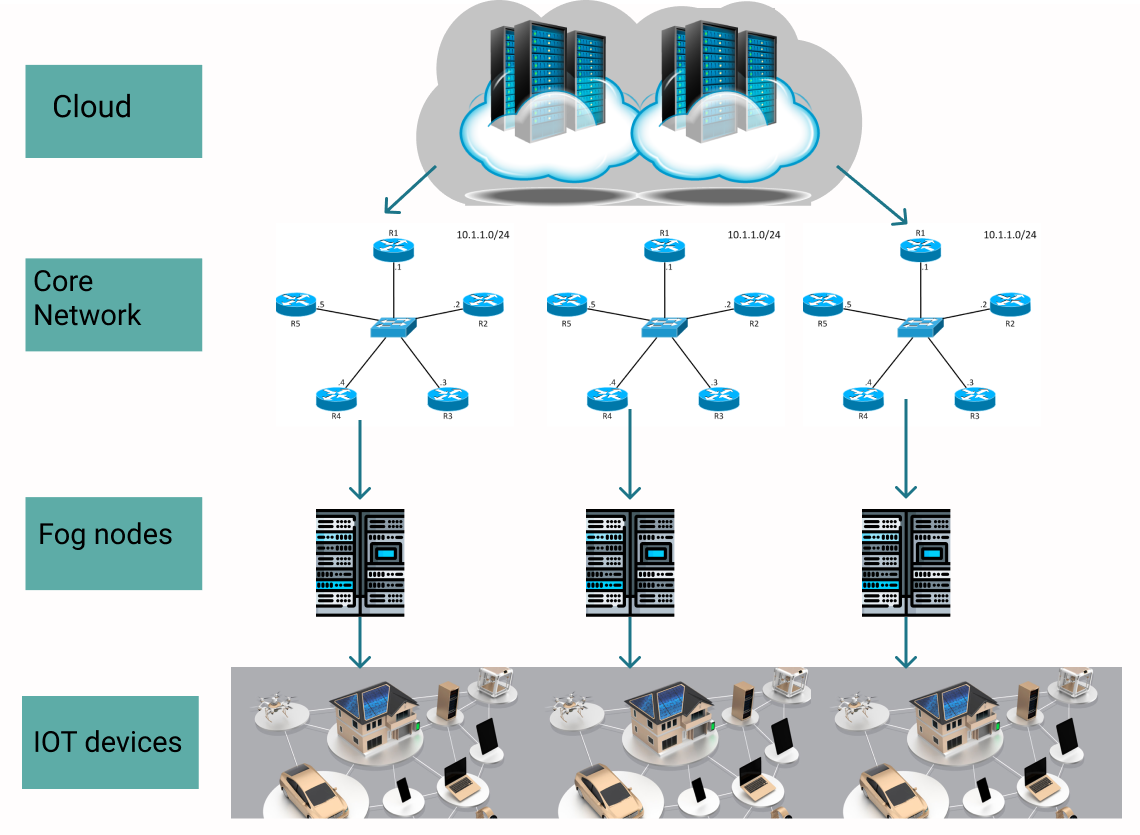}
\caption{A general model of fog systems}
\label{fig1}
\end{figure*}

As a general definition, fog computing is defined as a distributed computing architecture, in which we bring our storage and processing resources near the edge devices \cite{zhangfog}.  Fog nodes can be any nodes that have been equipped with communication, processing and storage capabilities. Clearly, the number of devices that are being connected to the internet is growing and the amount of bandwidth that needs to be provided for them is huge. Apart from bandwidth, there is a need for having a standard to translate communication protocols as well. Furthermore, there are other definitions for fog computing; Bonomi in \cite{bonomi2012fog} presents fog computing is a platform that has been virtualized highly to provide different services, such as computing, storage and network that is between cloud and IoT nodes. Its important to say that, they are not just installed near the edge of the network.\newline
Another definition is, fog computing can be defined as a scenario in which many heterogeneous and decentralized devices communicate with one another in order to do storage and computation tasks. There are some incentive mechanisms for the nodes that lease their capabilities for hosting services as well \cite{vaquero2014finding}. Zhang in \cite{zhangfog} defines fog computing as a decentralized architecture that the storage and computation resources are placed at the edge of the cloud, instead of making exclusive channels for cloud utilization and storage. Figure \ref{fig1}, has shown a general model about fog systems.

\subsection{Fog architecture}
\begin{figure}[htbp]
\includegraphics[width=0.49\textwidth]{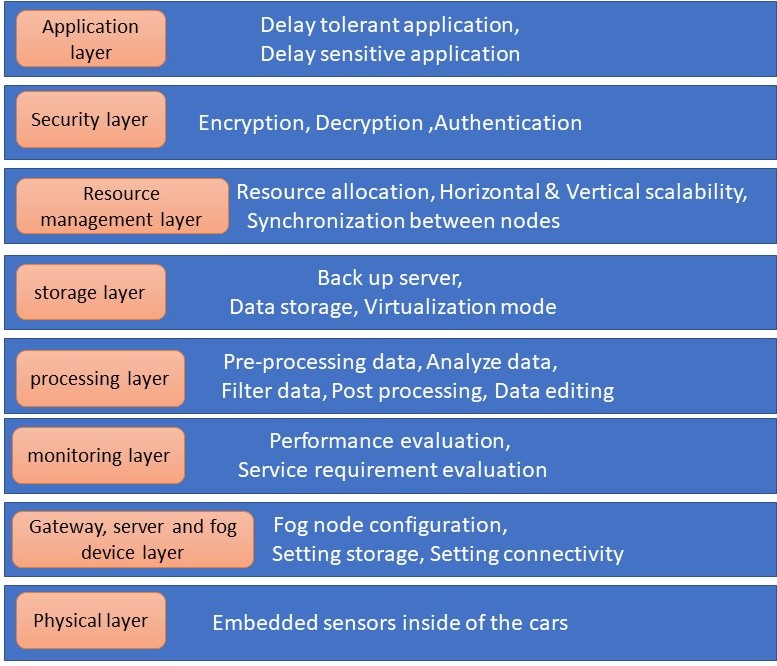}
\caption{8 layers fog architecture}
\label{fig3}
\end{figure}

There are various architectures with different layers that have been proposed for fog architecture \cite{dastjerdi2016fog}\cite{luan2015fog}\cite{aazam2015fog}\cite{sarkar2016theoretical}. There are three \cite{luan2015fog}\cite{mahmud2018cloud}\cite{buyya2009cloud}, four\cite{arkian2017mist}, five \cite{dastjerdi2016fog} and six \cite{aazam2015fog} layers in fog architecture. As shown in the figure \ref{fig3}, we can divide fog architecture into 8 layers \cite{naha2018fog}. 
\newline
\newline
\textbf{Physical layer:}
 This layer can generate the data. In the context of VANET, data is usually coming from embedded sensors in the cars, pedestrians, Road Side Units and cameras.
\newline
\newline
\textbf{Gateway, server and fog device layer:}
These entities can be an individual device or an IoT device \cite{taneja2016resource}\cite{giang2015developing}\cite{intharawijitr2016analysis}. However, as the name shows, fog server has higher capabilities compared to fog device. In the hierarchical manner, a number of sensors can be connected to a fog device like a car and similarly, a number of fog devices can be connected to a fog server. In some scenarios, fog server and different fog devices can run the huge tasks cooperatively. Moreover, this layer provides communication between fog servers.
\newline
\newline
\textbf{Monitoring layer:}
As it’s obvious, this layer is responsible for monitoring the overall performance of the network. It helps the fog device and server to decide which resources are better to use in different scenarios. This layer has the ability to not only evaluate the current resources' performance, but also it can predict the future needs for using resources based on the data load investigation and availability of the resources at the current time. Also, it has to meet SLA (Service Level Agreement) requirements as well.
\newline
\newline
\textbf{Processing layer:}
 This layer analyses data and data alteration would be performed when necessary. In this layer, we can decide to send the data to the cloud for further processing or keep it in the fog server for processing locally \cite{giang2015developing}. Furthermore, this layer has the ability to deal with the fault or missing data and reconstruct them after processing.
\newline
\newline
\textbf{Storage layer:}
This layer helps us to have storage virtualisation. Storage virtualisation is defined as a number of storage devices that can be connected together in the network and act as a single storage device. Moreover, this layer provides backup for data to avoid unwanted situations which take effect on the data.
\newline
\newline
\textbf{Resource management layer:}
Management and allocation of the resources, considering energy savings, occurs in this layer. This layer can ensure the reliability and scalability of the resources as well. Although cloud provides horizontal scalability, fog layer can provide both horizontal and vertical scalability. Another task of this layer is providing synchronization between multiple fog servers that are doing the same application simultaneously. Energy saving is performed in this layer to minimize the cost of using resources.
\newline
\newline
\textbf{Security layer:}
Like any other architecture, one of the important aspects in every architecture, is providing security. This layer provides security in different levels, such as: encryption and decryption of the messages, authentication, keeping the privacy of the fog nodes etc.
\newline
\newline
\textbf{Application layer:}
There are different types of applications that are delay sensitive. These applications can benefit from fog architecture as the computing and storage capabilities  near the edge devices. For example, real time traffic monitoring applications and all the applications that are based on Augmented Reality (AR), can benefit from fog architectures.
Figure \ref{fig3} has shown the 8 layers fog architecture model\cite{naha2018fog}

\subsection{Cloud}

\textbf{Cloud definition:}
Cloud computing idea has been defined by NIST \cite{mell2011nist} (The National Institute of Standards and Technology) as a model to provide unlimited network access to shared computing resources. There is a large pool of virtualized resources in the clouds that can be adjusted for different loads of work. Different adjustment for cloud services allows different users to have different accessibility to the resources of cloud data centres in the concept of 'pay as you go' model\cite{vaquero2008break}. Different customers can use desired resources based on their needs which are provided by large companies, such as Amazon, IBM, Microsoft, Google etc. Access level to the resources for each user is based on the amount of money that they pay for using resources.

\begin{center}

\begin{table*}[t]
\caption{difference between fog and cloud}
\label{diff-fog-cloud}
\centering
\begin{tabular}{ | c | c | c | } 
\hline
\textbf{Features} & \textbf{Fog} & \textbf{Cloud} \\ \hline 
Architecture & Decentralized/ Centralized & centralized    \\ \hline
storage \& processing nodes         & Any nodes that have storage and processing power & Powerful server type nodes        \\ \hline
Connectivity type                   & Highly wireless                                  & Combination of wired and wireless \\ \hline
Power use & Low & High \\ \hline
Computing \& storage capacities  & Low  & High           \\ \hline
Mobility of the nodes   & Highly mobile  & Low mobility   \\ \hline
Delay sensitive application support & Highly support & Low support \\ \hline
Location awareness  & Yes   & No     \\ \hline
Geo distribution& Distributed & Centralized    \\ \hline
Mobility support& Highly support& Low support    \\ \hline
Heterogeneity support& Highly support & Low support    \\ \hline
Internet access to get resources & Not necessary& Necessary \\ \hline
\end{tabular}
\end{table*}
\end{center}

\textbf{Cloud services:} 
There are three different types of cloud services IaaS (infrastructure as a service), PaaS (Platform as a service) and SaaS (Software as a service) that will be explained in the following sections. Also, all of these services are going into a service named XaaS (anything as a service)
\cite{evans2011internet}.
\paragraph{A)IaaS:}
\cite{dillon2010cloud}Infrastructure as a service is one of the service types that provides direct accessibility to the network, storage and processing infrastructures. For example, if someone wants to configure a Stand-alone VM for his company, he can directly arrange the number of requested CPU cores and RAM and other required infrastructures for his need through engaging with large companies that already offer their services to the customers.
\paragraph{B)PaaS:}

Platform as a service, which allows the customers to design and develop their own software through getting help from middle ware. This service is helpful for the people that do not need to set up the infrastructures and just need to manage their software life cycle. So, they can put their attention in developing their software, as required infrastructure are provided by the companies.
\paragraph{C)SaaS:}
Software as a service or SaaS is helpful when the customers do not want to deal with software challenges, such as managing databases, controlling socket and scalability issues. SaaS is used when a person doesn’t want to even install the software manually. In this regard, Google Apps can host client software as a web application.
\newline
\newline
\textbf{Cloud types:}
Apart from different types of cloud services there are four different cloud deployments: Private, Community, public and hybrid clouds \cite{mell2011nist}. \paragraph{A)Private clouds:} They are used just by a single entity and have high privacy. These types of clouds are completely independent of the model of 'Pay as you go'.
\paragraph{B)Community cloud:} In this type, the services and resources in the clouds are usually shared between the users from different companies in a decentralized controlling manner, as different organizations are using the different number of cloud resources.
\paragraph{C)Public cloud:} This type is offered by big companies, such as Google, IBM, Amazon, etc. and compared to private clouds, they are more user-friendly, cost-effective and easy to manage, and can take advantage of pay as you go model. 
\paragraph{D)Hybrid clouds:} As its name states, combine different capabilities from different cloud types \cite{sotomayor2009virtual}. 
Cloud computing was one of the popular ideas when it was proposed by the researchers. However, over time, it shows some drawbacks that cannot be practical for meeting delay sensitive and time critical applications, especially when the delay is so important for getting back the responses. At the same time, with deploying large number of IoT devices that each of them generates the large amount of data, a need for installing number of devices that are equipped with high storage and computation capabilities was identified. Finally, the idea of fog computing was established by academia and industry \cite{bonomi2012fog}.

\textbf{Difference between Cloud and Fog:}
As already mentioned in the definition of both cloud and fog computing paradigm, fog computing has a decentralized manner and provides computing resources near the edge devices. Having decentralized manner of fog computing allows the user to be a fog node, such as vehicles that can act as a fog node and host different requests from different users, or can be just a consumer node that benefits from computation and storage resources that are within its vicinity. Another major difference is about having different types of availability to the resources that are existing in the fogs and clouds. Fogs provide moderate availability of the resources to the consumers, while clouds offer high availability of them \cite{jalali2016fog}. In terms of size of the infrastructures, clouds use big data centres, whereas, fogs have small size infrastructures, such as small access points, gateways, routers etc. Another difference is, resources in clouds are accessible through network core, while, fog resources are accessible through devices that are already connected from edge to the network core. Another important distinction is, for using the resources in fog, the users do not need to have internet access all the time. In other words, having internet access to get fog resources is not essential, and resource access works with or without internet connectivity. The internet connection is needed for having access to fog resources in lack of other communication types, such as Wi-Fi. In \ref{diff-fog-cloud}, we summarized the main difference between cloud and fog, such as mobility of the nodes or location awareness. In the context of VANET, cars can act as a fog node as well.
\section{Preliminary terms}
\label{pre-term}
\subsection{What is 5G}
5G is defined as a next generation of mobile broadband which has higher download and upload speed with features that differentiate 5G from current cellular communications, such as 4G LTE. It helps the devices to connect to wireless networks with the least amount of time.
\newline
\newline
\textbf{Different spectrums in 5G:}
5G is different in some aspects compared to current cellular networks. The differences between different spectrums in 5G will be explained in the following paragraphs:

\subparagraph{1)low band spectrum:}
This spectrum was a primary spectrum that is used by some carriers in the US and can be defined as a sub 1GHZ spectrum. It has some benefits, such as: good coverage, and has the ability of wall penetration. On the other hand, it has some disadvantages, such as: 100 Mbps upper limit for internet speed. In low-band spectrum, T-Mobile plays a really important role to increase a massive amount of 600 MHz spectrum at FCC (short form of Federal Communications Commission).

\subparagraph{2)Mid band spectrum:}
Compared to low band spectrum, Mid band spectrum provides lower latency and faster speed. In terms of drawbacks, it cannot penetrate the wall like low band spectrum. We expect the speed to reach to 1Gbps in this spectrum. Some carriers in the US have the large number of unused mid band spectrum, and they use Massive MIMO in order to improve penetration and increase coverage area in this spectrum. Massive MIMO takes a group of antennas and puts all of them into a box on a single cell tower in order to establish simultaneous beams for multiple users. Another technique which is used on the mid band spectrum is using beam forming technique by some carriers. In beam forming, we have a focused signal that is sent to every single user, and we monitor continuously to make sure they get a signal that is consistent.

\subparagraph{3)High band spectrum:}
 This type of spectrum is usually called, mm Wave and shows the best performance that is existence in 5G, although it has some minor weak points. It offers 10Gbps speed and, it has the lowest latency compared to other spectrum band types. As mentioned above, it has a few drawbacks, such as low coverage and poor wall penetration. It goes without saying that, as the coverage area is minuscule, we need thousands of cells in our network. So, some carriers are using a large number of small cells to cover the weaknesses of high band spectrum and at the same time they use beam forming technique to make sure each node receives a constant signal.
 \newline
\newline
\textbf{5G requirements:}
There is an agency named ITU (International Telecommunication Union) that establishes and develops required standards for different new communication technologies. In other words, it is responsible for creating rules in the usage of radio spectrum and telecommunications. In order to research and develop the minimum needs for using 5G, a program named IMT-2020 was created by ITU in 2012. In 2017, thirteen minimum requirements were established by IMT-2020. At the same time, 3GPP (third Generation Partnership Project) started working on some standards for 5G. Finally, they established two standards: non-standalone (NSA) specification which was released in 2017 and stand-alone specification (SA) which was released in June 2018. Both NSA and SA have the same specification, with one major difference. NSA exploits current LTE and SA uses core network of a next generation for roll-out. 
3GPP standards relatively corresponds with the targets of IMT-2020 in terms of performance. They are listed in here:
\paragraph{Peak data rate:}  In 5G, we have high speed data communication. The speed of down link is 20Gbps and the speed of up link is 10Gbps for each mobile BS. Its important to note that, this speed is not a dedicated speed for one user, but it’s the speed that is being shared between all the users in one cell.
\paragraph{Real World Speeds:} It’s interesting that the highest data rate in reality is completely different from the mentioned peak data rate for 5G. There is 100 Mbps for download and 50 Mbps for upload for each user.
\paragraph{Latency:} As mentioned earlier, 5G provides the lowest latency compared to other cellular communications. There are 4 milliseconds latency for data travelling from source to destination in the normal situation. Moreover, in time sensitive applications or real time data that needs the lowest latency, there is 1 millisecond latency.
\paragraph{Efficiency:} One of the noticeable things in 5G is switching between different radio interfaces based on whether they are in use or not. In other words, radio interfaces should have the ability to change into low energy state when they are no longer in use in less than 10 milliseconds.
\paragraph{Spectral efficiency:} It's important to  optimize the use of each spectrum, or even using the bandwidth which leads to sending maximum data with minimum transmission errors. 5G needs to have spectral efficiency as well that is slightly over LTE, having 30bits/HZ for down link and 15 bits/Hz for up link.
\paragraph{Mobility} Although mobility management in LTE is easy, it's challenging in 5G, as we have millimetre wave networks. Supporting movements from 0 to 310 miles per hour in every base station in 5G is necessary. That means base stations should have the capability to work with a range of antennas that are moving.
\paragraph{Connection density:} Compared to LTE, 5G needs to support more devices per square kilometre. The number of devices to support is about one million which is a big number compared to 4G LTE, that shows we are powering the IoT (Internet of Things).
\newline
\newline
\textbf{5G in usage:}

\emph{Healthcare:}
One of the components in the context of 5G named URLLC (short form of Ultra-reliable low latency communication) can fundamentally improve and change health care, as it decreases the response time to the patients even further than current cellular communication and its related applications.
\paragraph{IoT:}
 One of the crucial features of 5G is supporting huge number of devices in the network that all have access to internet communications. In 4G LTE, as the number of offered resources are not enough, communication between the devices encounters many problems, and finally, we reach the maximum capacity of LTE. Through smart devices, such as massive Machine Type Communication (mMTC) devices, there is no need for huge resources, because they can connect to one BS that is much more efficient.
\paragraph{Improved broadband:}
Undoubtedly, one of the important things that 5G can do is improve bandwidth capacity that current cellular communication is unable to provide, especially during high traffic periods. 5G can exploit using spectrums that have not been used before, especially for commercial broadband traffic.

\paragraph{Public safety infrastructure:}
 5G can help governments and other companies in emergencies, such as flooding or earthquakes through using sensors that are implemented in various cases to work together and notify people about the events.

\paragraph{Vehicular Networks:}

It can be said that communication between cars without using 5G technology is extremely difficult, especially when the real-time data analysis and time-sensitive applications come into play. The communication between the cars and infrastructures can be defined as V2V and V2I. Suppose a car that wants to inform other vehicles about collision in less than 1 millisecond. To do that, it needs to take advantage of 5G that provides this capability.

{Communication types in Vehicular Networks:} 
\begin{itemize}
    \item V2V: It’s defined between cars, and they can inform each other about different events and information on the roads, such as collision, traffic status, safety collision, etc.
    \item  V2I: Occurs between vehicles and infrastructures. The infrastructures can be anything that are along the roads, such as traffic lights, bus stops or even some buildings equipped with different types of communications.
    \item V2P: It is run between vehicles and pedestrians. In this communication the cars can inform passengers about the pavement, some events in some locations, parking spots, etc.
    \item V2N: It is the communication between vehicles and internet. The cars can get the required information through cellular communication when they are in the road. Its important to note that, in some papers V2N can be in the communication field of V2I that support cellular communications as well.
\end{itemize}

\subsection{Software Defined Network (SDN) in Vehicular Network}

The concept of SDN is simple and straightforward in vehicular networks, as they helped us to separate the control plane and data plane and easily manage our network that can satisfy the current challenges of limited resources. There are many research papers about the integration of SDN and VANET. However, because of the dynamic nature of VANET, adopting SDN with VANET is challenging in many cases.
However, centralized control of SDN management can solve many issues that we have in VANET, such as high mobility of the nodes.Also, advantages of SDN in VANET can be categorized into three parts:

\paragraph{A)Routing and spectrum management optimization:}
It can help the vehicles to send their packets properly in some places that are overcrowded through having a good policy for routing packets to decrease congestion. Moreover, it can optimize using spectrum. SDN can decide to use the best channels in the case of an emergency.

In the context of IoT, the management of the spectrum seems crucial. There is a need to have a dynamic and flexible approach to manage spectrum resources. Connecting IoT through 5G technologies needs a large amount of spectrum resources, coupled with managing a huge number of connections and security. There is a term named network densification that is used when we deploy a huge amount of wireless infrastructure \cite{luan2016guest}. 5G technology needs D2D communication with having a bigger coverage area to support proximity-based services that provide a larger coverage area. Also, we have an intensive radio coverage in  VANET, and many vehicles that are in the vicinity of each other, can access the network resources at the same time. However, the number of vehicles that have access to these resources has been limited because of limited spectrum resources and having low spectral efficiency. In this way, authors in \cite{huang2017exploring} designed a system named VNG (short form of Vehicle Neighbor Groups) that provides the opportunity that a huge amount of data packets from connected vehicles are delivered simultaneously, when they are in the same location. It would be more interesting and helpful when the neighbour vehicles have the same destination or the same path and can share their experience in network access.

\paragraph{B)Optimizing resource management in the network:} 
SDN can help the network to optimize using resource management, SDN can decide to adjust transmit power, or not, to increase the chance of packet delivery in the network. However, in VANET environment, control plane in SDN has some challenges in optimizing resource management of multiple RSUs. In other words, when the vehicles move into the overlapping coverage area of two RSUs, it is challenging for the SDN controller to decide which RSU it should connect to. Furthermore, when the large number of cars are just connected to one RSU, frequent switching challenges degrade SDN performance as stated in \cite{ghafoor2012anchor}. There are some solutions to resolve the problems and to make the performance of SDN better in VANET \cite{duan2016sdn}\cite{li2016control}, such as considering latency requirements and cost of cellular networks or decreasing bandwidth usage through sending SDN control request by vehicles in the cellular network. Duan et al.(2016) in \cite{duan2016sdn}, proposed a new architecture to combine SDN and CRAN to fully utilize the flexibility and centralization capabilities of SDN. This new architecture proves that regardless of how dense our network is, they can provide minimum delay compared to traditional approaches. There are other infrastructures, such as combination of 5G and SDN for having a secure design in VANET \cite{hussein2017sdn}.
\paragraph{C)Energy management in 5g in VANET:}
As it's obvious, when we have network densification, ultra-dense deployment in the network leads to huge energy costs that should be managed to decrease the costs. In some studies, researchers concluded that many BSs are not exploited completely most of the time, and using them is completely dependent on the time and location. Although the cars are not using them most of the time, for example at weekends, they use most of their energy (90 percent approximately). It goes without saying that, conserving energy is one of the main necessities that should be deeply studied. There are some approaches for energy management that are outside the scope of this paper, such as proposing NOMM (nonorthogonal multiplexing modulation) in order to collect V2I traffic to improve energy efficiency \cite{niu2010cell}. Another approach is monitoring the network and activating the base stations that satisfy our needs and turning off the others in order to save the energy. Also, in order to get the minimum energy consumption, modifying the coverage area in each cell based on the traffic load changes is considered as another promising technique as well.  \cite{niu2010cell}.

\subsection{Other Related Technologies} 

\textbf{What is F-RAN and C-RAN?}

 Putting the ideas of both fog and mobile technologies together, creates a new term named Fog RAN (Radio Access Network) which leads to faster data access in the edge, and it's possible to implement through 5G technologies\cite{hung2015architecture}. Another term is named cloud RAN which benefits from virtualization, and has a centralized controller over all the F-RAN. There are many radio heads which are remotely and randomly implemented. Through front haul links, they are connected to the BBU (short form of Base Band Unit) in C-RAN. Moreover, C-RAN is also performed in 5G technologies concepts. As mentioned in \cite{peng2016fog}, both of these technologies provide a form of energy-efficient network.

\textbf{Mobile computing:}
When computing is run on the mobile entities, we name this type of computing, mobile computing. Mobile computing devices can be mobile phones, laptops, cars, and other mobile devices. There are some challenges in the context of mobile computing, such as intermittent connectivity between nodes, low bandwidth, having a heterogeneous nature of mobile devices, etc. One of the solutions is robust caching that is proposed in \cite{forman1994challenges}. Because of the different challenges in mobile computing, this type is not adequate for different needs of consumers. So, there is a need to combine both concepts of fog and cloud to cover the shortcomings of each one with the advantages of the other to extend the scope of mobile computing to be independent of the local network. The difference between mobile computing with fog and cloud computing is, in mobile computing, we have limited resources that are not comparable with fog and cloud in terms of computation and storage power. Moreover, there is a need for the users to adapt themselves in different environments. Furthermore, although this type of computing can benefit from distributed computing, time sensitive applications, large data processing, and storage requests cannot take advantage of this computing \cite{satyanarayanan1996fundamental}. On the other hand, the combination of cloud and mobile computing can create a new term named mobile cloud computing (MCC). In MCC, the capabilities of storage and computation are placed on the outside of the mobile devices \cite{dinh2013survey}. There are some applications in MCC, such as healthcare, image processing sensors, task offloading, and crowdsourcing that are explained fully in \cite{ren2015exploiting}\cite{sanaei2013heterogeneity}. Task offloading can help the mobile devices to not only offload some parts of their huge tasks to the fogs or clouds to finish the tasks at a right time \cite{shiraz2012review}, but it helps them to save their battery life.  Also, security issues are vital to take into consideration. There are some drawbacks regarding MCC, such as it is not adaptable for the applications that need pervasiveness of the devices, as computation and storage have occurred in the cloud.  Besides, when the devices want to offload their huge task to the cloud, other issues arise, such as latency and security. 
\newline
\newline
\textbf{Mobile ad hoc cloud computing:}Mobile ad hoc cloud computing is the most decentralized type of architecture. It creates the clouds in the ad hoc network to help the system to have better storage and computation. Autonomous vehicles, streaming a live video in a group, and informing users about live disasters and managing the disaster in different strategies are a few examples of Mobile ad hoc cloud computing. It is worthwhile to mention the difference between MACC and MCC. In MACC, the location of computation and storage of data occurs in the mobile devices, and forms a dynamic cloud, while in the MCC it is executed far from mobile devices and occurs inside the cloud. Furthermore, compared to fog computing, MACC is more decentralized and more suitable for the applications that have a decentralized manner and desire pervasiveness of the devices.
\newline
\newline
\textbf{Edge computing:} 
Edge computing is one of the computing types that is done near the IoT devices, not inside the IoT devices. It can be one hop or two hops away from IoT nodes. OpenEdge computing \cite{gupta2019emerging} describes edge computing as the small data centres near the IoT devices that are responsible for data computation and storage. As edge computing is in the vicinity of the users, it can be easily dealt with privacy, intermittent connectivity, and delay-sensitive applications. It’s worth noting that, we have a lower latency in edge computing compared to MCC and cloud computing, mainly because of having adequate resources near the IoT devices. In terms of service availability between edge, MACC and cloud computing, edge computing has a higher service availability, because users do not need to wait for the cloud computing to provide the requested services for them and compared to the limited available resources in MACC, resource availability in edge computing is higher. it's important to note that some papers consider edge computing and fog computing the same. However, Openfog Consortium clarifies the differences between them. In fog computing, we have a hierarchical behavior. In other words, computation, storage, and other capabilities that we put them into the fog are distributed from cloud to the things. On the other hand, we do not have this facility in edge computing, as they are fully deployed in the proximity of IoT devices that might be one or two hops away from them \cite{openfog2017openfog}. So, when we talk about an edge, the location of the edge is one hop away from the IoT devices (like access points). There is another term named mist computing, which is used when the computation is run inside the devices.
\newline
\newline
\textbf{Mobile Edge computing:}
Mobile edge computing is defined as another branch of edge computing, in which it develops edge computing by preparing resources of computation and storage in the proximity of mobile devices that have limited sources and life battery. The common point between edge computing and mobile edge computing is, both of them can be run from the places, or edges that have full internet connectivity and can be extendable to the places that have no internet connection. The difference between them is edge computing only arranges connections through WiFi, WAN, and cellular connections, while, MEC can create any type of connectivity.  Moreover, it can be said that MEC can help edge computing to get access to a whole range of mobile users with minimum latency and manageable core networks that are mobile \cite{taleb2017multi}. Time-sensitive applications can be supported by MEC as well \cite{hu2015mobile}.
\newline
\newline
\textbf{Cloudlet computing:}Cloudlet computing was created to cover some drawbacks of MCC. A cloudlet is a group of computers that have enough resources of computation and storage and all of these computers inside the group have access to the internet connection where mobile nodes can take advantage of them \cite{satyanarayanan2009case}. Clearly, they are small clouds with  enough resources, and they are one hop away from nodes. When the mobile devices are unable to compute their huge tasks, they offload their tasks to the cloudlets that are located one hop away from them \cite{hao2017edgecourier}. Having the advantages of cloudlets, different providers can provide cloudlets for their users and put these small data centers on the network edge. The difference between cloudlet and MACC is, cloudlet has an infrastructure that has been virtualized and is in the form of Virtual Machine (VM), on the other hand, MACC suffers from the lack of virtualization of devices especially for time-sensitive or real-time analysis.
\newline
\newline
\newline
\textbf{Mist computing:}Mist computing is defined as computation inside the IoT devices\cite{preden2015benefits}. Mist computing is the first location of the computation and evaluation of whether the tasks can be performable inside the IoT devices or not. 
\newline
\newline
\textbf{What is vehicular cloud computing:} 
There are many vehicles with having enough computation and storage power on the street that can be taken into consideration as the potential computation and storage resources for data analysis. Also, many cars spend most of their time in the parking spot without exploiting their resources. So, The features of vehicles transform them into a perfect candidate for a cloud computing network.Moreover, to convince the drivers to share their car’s resources, many incentive mechanisms have been proposed to motivate them to renting out their car’s computation capabilities, such as using parking spaces freely. The idea of vehicular cloud computing can be defined as a cluster of vehicles that can share their computation and storage resources to serve their high demanding tasks in real-time or non-real-time applications. Vehicular cloud computing can be formed through several vehicles that have been parked in a parking area without having mobility, or can be formed through some moving vehicles that are in the proximity of each other on the road. Vehicular Cloud Computing is useful in many time-critical applications, such as traffic data analysis.
As this type of cloud plays a really important role in real-time applications especially in the case of disasters, the Federal Communication Commissions in the USA allocated a specific spectrum for supporting communication between vehicular networks named: DSRC (short form of Dedicated Short Range Communication) with the range of 75MHZ.
\newline
\newline
\textbf{Vehicular cloud computing architecture:}It has 3 different layers. The first layer is the vehicle layer. This layer is usually equipped with OBU (Onboard Unit) and they can be different types such as steam, pollution, temperature, speed detection sensors and so on. They can collect all the related data and then send aggregated data to the second layer. Also, OBUs are equipped with different types of communication, such as cellular and DSRC. Second layer named communication layer in which, there are two types of communications named: V2V (Vehicle to Vehicle communication) and V2I (Vehicle to Infrastructure) that can be supported over satellite,DSRC, and cellular communications. The last layer of this architecture, named cloud computing, which has the whole data of the network that has been aggregated by OBUs. This data is useful especially when some organizations such as hospitals, police stations, etc. want to fully exploit them for their specific purposes. Cloud layer itself, consists of application sub-layer, cloud infrastructure, and cloud platform. Application layers can be real-time applications and are accessible through vehicles, and data that are gathered from cars, can be used by these applications. There are different services in this layer such as: NaaS (network as a service), Storage as a Service (STaaS), Cooperation as a Service (CaaS), Entertainment as a Service (ENaaS) and finally Information as a Service (INaaS).
\section {Vehicular Social Networks}
\label{VSN}

The combination of two different areas, namely Vehicular ad hoc network and mobile social network, makes a new paradigm which is Vehicular Social network (VSN). In VSN, each vehicle shares its data with the other vehicles that have similar interest in its vicinity \cite{vegni2015survey}. Based on social behaviors and common interests among the cars, each car can select other cars as its friends. In this section, different features that are adaptable in VSN, and differences between VSNs and Online Social Networks are discussed. After, we discuss current challenges in VSN. For better understanding of common interests between vehicles to make a group of friends, Xie et al.(2014) \cite{xia2014beeinfo} proposed a routing system named (BEEINFO) that group communities in different groups based on their similar interests. The main idea of this system works on recording and getting the information from passing different communities by mobile users (driving cars). If the number of nodes in a community is large, it shows that the number of members in that group is also large. Imagine three different groups are related to these places: shopping mall, school, and hospital.  If two vehicles pass from shopping malls and hospital, and the density of one vehicle in one group is higher than the other ones (for example one of them is a bus and the other one is just a car), the vehicle that has higher density has higher chances to be a forwarder for the messages \cite{bradai2014reviv}. Moreover, within each community, one or more nodes that have higher social ties with other nodes can be selected as forwarders.
In vehicular social networks, there are two important layers which are physical layer and social network layer that cooperate with each other. For more clarification about the social behavior of vehicles, in \cite{srivastava2014social}, they discussed SNA (Social Network Analysis). In this analysis, they considered different metrics that are important in the context of social relationships among vehicles. A social network is composed of social ties between the users and their common interests. It can be said that, all these parts are important in terms of affecting the structure of social networks \cite{wang2013fine}. In VANET network, there is always a need to check the connection between the nodes and check the topology of the network, as they constantly change over time because of the dynamic nature of VANET. In this sense, SNA checks and monitors the traffic topology over the day to perceive the human behavior coupled with rush hours, and similar trajectories \cite{bradai2014reviv}. As mentioned above, three different metrics in SNA are important, named degree centrality, betweenness centrality, closeness centrality, and bridging centrality. 

\subsection {Different Metrics in SNA}

\begin{figure}[htbp]
\includegraphics[width=13cm, height=7cm]{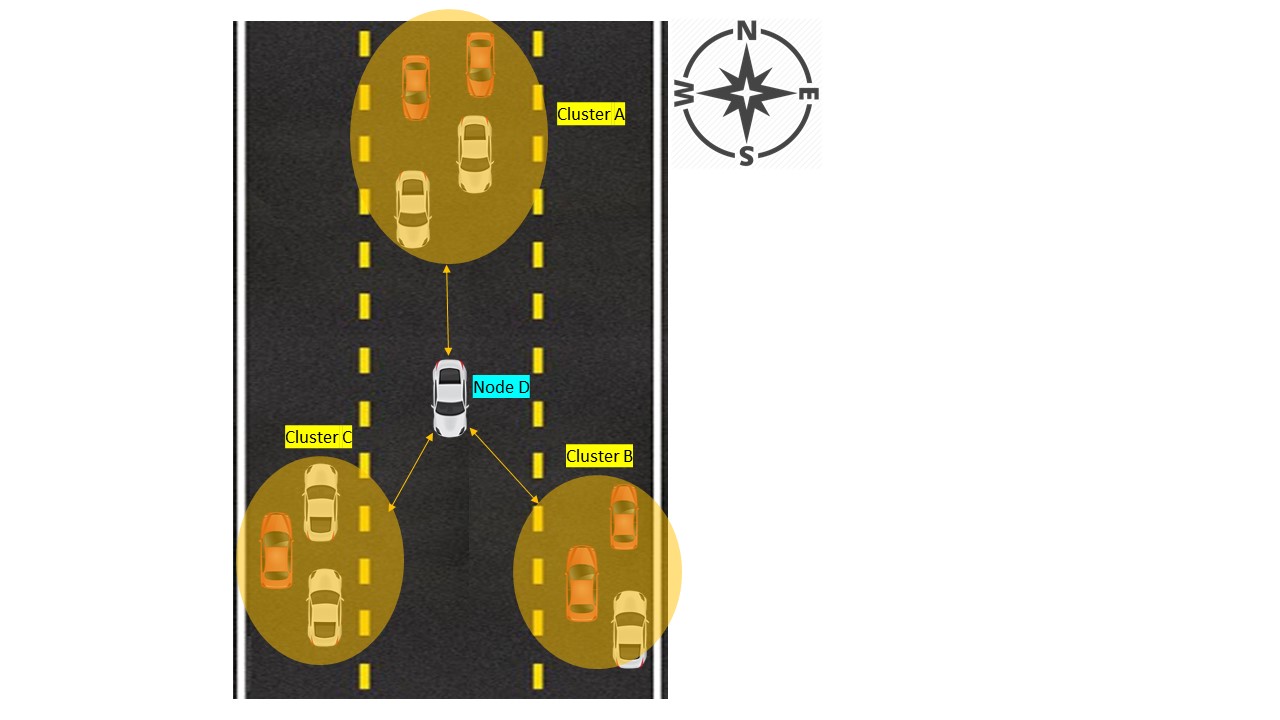}
\caption{Node D has the highest centralities metrics}
\label{high-node-centr}
\end{figure}

Social Network Analysis comprises a number of metrics in order to optimize the network. They have been briefly explained below.
 \paragraph{Degree centrality:}  A node is considered as a high degree centrality when it has the highest number of connection with neighbouring vehicles. In other words, when the number of nodes that are communicating with one node is high, that node has higher degree centrality compared to the other nodes and has a higher chance to forward the messages to the other nodes successfully. Thus, when a node is about to send packets to other nodes, it selects a node that has higher social or degree centrality as a forwarder in order to make sure that those packets will be delivered to the destination successfully.
\paragraph{Betweenness Centrality:} Having a node with the highest betweenness centrality is important for the network in terms of connectivity between the clusters, as having higher betweenness centrality for one node (node D in figure \ref{high-node-centr}) helps the network for having a greater number of the shortest paths among the clusters. 

\paragraph{Closeness centrality:} It means how much a central node is close to the other nodes that are in its proximity. 
\paragraph{Bridging Centrality:}  a node with high bridging centrality is defined as a node that has the highest connection and shortest path between different clusters in VANET. In the formula, there is one metric named bridging coefficient that identifies how well a node is placed between a high density of nodes.

For better understanding, figure \ref{high-node-centr}, shows 3 different clusters, A, B, and C. A is a cluster of vehicles that are going from north to south along the road. On the other hand, B and C are other clusters that are moving from south to north. At its clear, all of these clusters have the chance to connect to node D who has the highest betweenness, closeness, and bridging centralities. Without node D, these 3 different clusters have a low chance to connect to each other, and that’s why node D acts as a relay node. Furthermore, pair nodes are connected together through at least one short path. 

There are different metrics for selecting a node as a relay node. Gu et al.(2014) in  \cite{gu2014social} showed a method for forwarding data packets through different metrics. A node is chosen as a next-hop to not only follow the rules of greedy algorithms, in which it should have the shortest distance to the destination, but also it needs to consider social features, such as high centrality. Furthermore, smilovic et al.(2012). in \cite{smailovic2012bfriend}, established a social graph in ad hoc network through inspiring Facebook and existing social graphs in that.

\subsection{Social Features in Vehicular Social Networks} 
 There are different factors related to human behaviours, such as their mobility, different preferences, selfishness, and some other factors that can largely affect the topology of a vehicular network. In \cite{cunha2013effective}, the authors discuss the behaviour of vehicular networks in the context of social manners. Through some numeric analysis, they concluded that the social behaviours of the cars definitely affect the performance of the network. So, considering social manners in the network topology, can improve its performance in terms of multiple services, such as safety and non-safety applications and communication protocols. In addition, through considering the social behaviour of the cars, we can avoid the broadcast storm problem in the network. In other words, only nodes that have a higher chance to deliver the packets successfully, are chosen as relay nodes (based on their social behaviour) which leads to avoiding multiple times forwarding in the network. Maglaras et al.(2015) in \cite{maglaras2015social} proposed a method named SPC (short form of Sociological Pattern Clustering) that considers the social manner of the vehicles and their preferences to share the same routes.
There are three different factors in human behaviors \cite{bradai2014reviv}: 1) person preferences 2) her selfish status 3) person mobility model. For example, people can make different decisions in different scenarios based on their preferences. In rush hours, some of them prefer to choose the longest path that doesn’t have traffic to go to their destination and some of them prefer to choose the shortest path (in terms of distance) to get to their destination regardless of having traffic in that chosen path or not. 
In \cite{lu2013bounds}\cite{lu2012capacity}, they proposed a model that shows a mobility model that identifies some social spots such as hospitals, malls, cinemas. They put these social spots in a gird that is scalable, and the mobility model of each car is restricted to a social spot. There are some mobility patterns such as IMPORTANT \cite{bai2003important}, GEMM \cite{feeley2004realistic} and BONNMOTION \cite{ai2014social} for vehicular social networks. As mentioned above, selecting a node as next hop for forwarding packets depends on different factors and one of them is the selfishness of nodes. This behaviour can affect the performance of the network. Because of some reasons, some nodes are not really interested in packets forwarding or sharing their computational and storage resources. So, it's important to find selfish nodes and evaluate their behaviour whenever we want to choose a node as a forwarder. One of the schemes in this regard is reputation criterion \cite{bigwood2011ironman}\cite{xu2013cooperation}. In this method, a node is selected as a forwarder based on its reputation history in which whenever it received a packet, it forwards it to the related destination without dropping that packet. This node is considered a good behaving node and if it didn’t have a good reputation in forwarding packets such as dropping the packets several times, it is considered as misbehaving nodes, and it won’t be a part of packet forwarding in the network. Another method for managing selfishness nodes named TFT (short form of Tit For Tat) \cite{zhou2014incentive}, in which every node forwards messages to its neighbours depending on how many packets it has received from its neighbours. This method is working based on the behaviour of the nodes that do not have a good reputation in forwarding packets. In other words, if a node drops a packet that is related to its neighbours, the next time its neighbours do not send its packets to misbehaving nodes. Another method named SCR (short form of Social Contribution-based Routing) \cite{gong2014social} in which they put emphasis on how social contribution can be an important factor for the cars that act as a selfish node and how much this metric can encourage them to be a part of forwarding packets schemes. For more clarification, they took two metrics for selecting relay nodes: 1) a node that has a higher probability for delivering the packet and at the same time 2) a node that has the highest social contribution in the network. 
In addition, there is a term named Social Internet of Vehicles (SIoV) in the context of IoT, that can combine the idea of social networking in VANET into IoT. SIoV can provide data for different services throughout the network, such as road safety application or managing traffic in rush hours. It's important to note that there is another term named Social Internet of Things (SIoT) \cite{atzori2012social} in which every node is capable of creating social connections with other nodes automatically. In SIoV there are different social connection between nodes that is :
\paragraph{POR(Parental Object Relationship):} It has been defined between the same manufacturer that provides helpful data about car status. 

\paragraph{SOR(Social Object Relationship):} This relationship considers the same locations and common routes that happened among vehicles. In other words, its created between cars through V2V links.

\paragraph{CWOR (Co-Work Object Relationship):} It's created between V2I (vehicle to infrastructure) that considers the relationships between Road Side Units and vehicles that continuously are in touch with together. This relation can be useful in some cases when the cars want to inform about traffic status and choosing the best route to get to their destinations. Moreover, it can be said that the daily routine of the cars can be predictable on the roads by monitoring them continuously to see what mobility pattern they follow every day, from their source that can be their homes to their destination that can be their offices. Based on their context-awareness in \cite{yasar2010people}, the authors developed a system named the Ubiquitous Help System (UHS). In this system, only relevant data that is important and useful between cars can be shared between them based on the social connection between them (Friend of a Friend or (FOAF)). 

\subsection{Vehicular Social Networks versus Online Social Networks}

There are some critical differences between VSNs and OSNs. Firstly, the social connections between vehicles happen online in which if two vehicles meet together, they can make a social relationship and these social ties can be strong or weak based on the different times that they meet on the roads or their common interests and destinations \cite{mohaien2013secure}. On the other hand, there is an offline social connection in online social networks, in which, people of a social network can be members of social interactions like friendship. There are different social webs, such as Facebook or YouTube, and based on a similar interest in some data that have been shared, people can establish social connections together. On the other hand, this issue does not happen in VSN. As there is a dynamic behaviour in the nature of VSN, social connections between members of a VSN can only happen due to encounters on their routes. Thus, one of the most important puzzle pieces in establishing a social connection between different nodes in VSN is the position of a vehicle. Suppose a car is moving along the roads that have numerous opportunities to join the different social groups. It can decide to leave the current social group that it is already in and join the other social groups. It is important to note that sometimes a car just meets one social group once and never meets its members again, and all of these issues are basically related to the position of the vehicles. Another important factor for being a group member in VSN is about how much the vehicles need some necessary content such as traffic status. Suppose that there are different social groups that talk about different criteria, such as traffic, weather condition, online gaming etc. Based on vehicle requirement, vehicles can join one of these groups that discuss relevant content together. Another metric that establishes a social connection between vehicles, is about having common interests, such as school workers, football attendants, football matches, and so on. The members of these social groups are the nodes with common interests. All in all, through different social networks based on these three different metrics (position, content, and relationship (or common interests)), there are different types of social networks on the road when the vehicle moves from point A to point D. Furthermore, apart from all differences between OSN and VSN, the same issue that is current in both VSN and OSN is about security and different threats that can affect the performance of the whole network \cite{fire2014online}. One of the approaches for joining into the suitable social groups in VSN is having a centralized approach that is performable through V2I connections. In other words, suppose that RSUs in a specific region has all the information about different social networks in that region, such as information of Museum social network, Churches social networks, and so on. After the vehicle gets information about all different social groups, based on its needs, it takes a decision to join to one of them.
Another approach named distributed approach that is happened through V2V communication protocols. This V2V communication occurs between the nodes that are inside the social networks. For example, if they belong to the social network named museum info group, they exchange the data among themselves to figure out what museums are open and which one is the best option to visit. It’s important to note that, all of these social connections are being built on the fly, that means they are completely connected to the specific locations and specific times. For instance, if an accident happens along the road, there is a social network in that road named accident info social network that members in this group exchange data about the severity of the accident, current traffic status and so on, which are completely related to certain times and certain area. Because of the short life of each social network, we can say that we have a “sporadic social network” \cite{bravo2013leveraging} in VSN.

\section{Related works}
\label{rel-work}

This section investigates the existing works in the literature focused mainly on recent advancement in cloud and fog architectures, Vehicular Social Networks, and the exploitation of 5G technology and Software Defined Networks to resolving current challenges exist in VANET networks. After, we compare recent works through different metrics with together. 

\subsection{ Recent architectures in the scope of Cloud and Fog in VANET}

This section discusses various hybrid architectures of cloud and fog in the scope of VANET. Then, to evaluate how much they have covered important metrics that is needed in fog and vehicular social networks, comparison in different metrics will be provided in a couple of tables.

\begin{figure}[htbp]
\includegraphics[width=10cm, height=7cm]{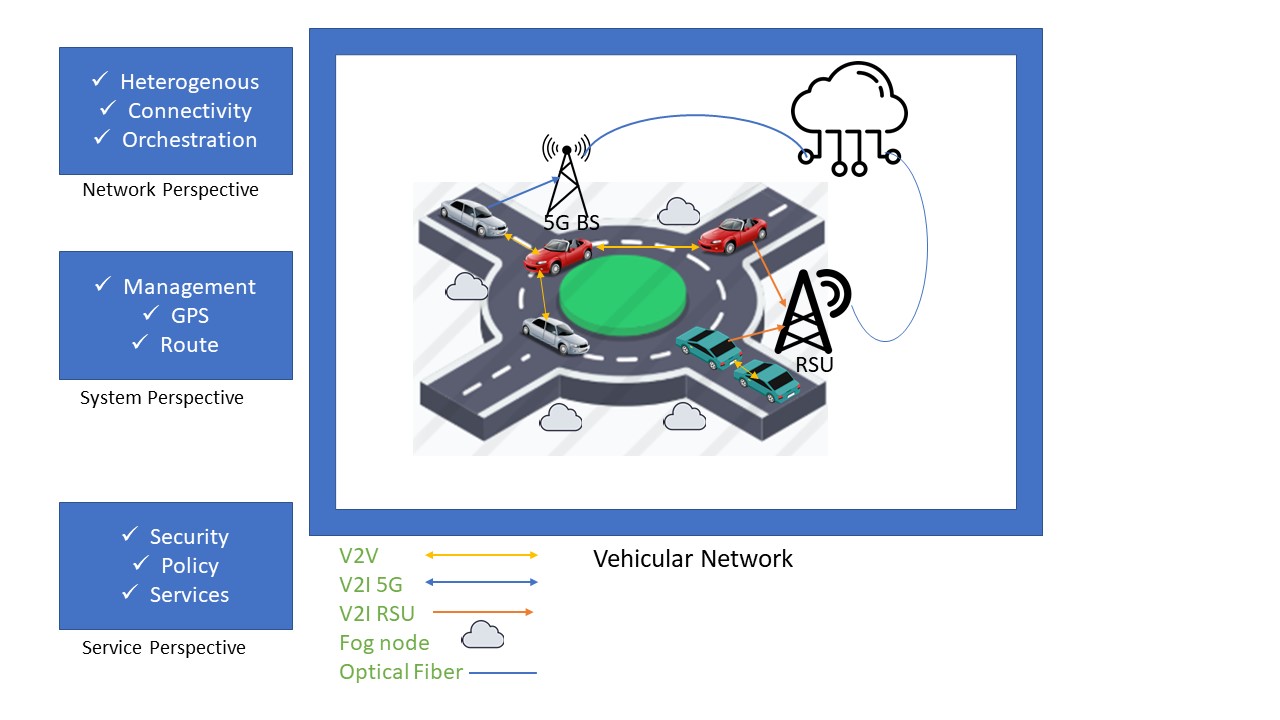}
\caption{Having System,Network and Service perspectives in VANET Scenario}
\label{picture1}
\end{figure}

In \cite{nobre2019vehicular}, they have used SDN together with fog nodes to achieve higher efficiency through using real data in Sao Paolo of Brazil. Then, they concluded that we can have minimum latency to be present at the accident location with a combination of fog computing and Vehicular Software Defined Network (VSDN).SDN gives the users a clear-cut management of data as it splits the network into 2 planes: control and data planes. Data plane as its name shows, is responsible for gathering data and sending the data in a specified location. Control Plane manages and controls the data in a centralized manner. In the idea of VSDN, we have different vehicular nodes that are being managed by a centralized controller in cloud computing. They consider  challenges in three different views: System, Service, and Networking. In the context of Networking, because of the heterogeneity of the network, we have different types of communication such as V2V and V2I. For dealing with intermittent connectivity in the network perspective, we have to update the network topology by periodically broadcasting messages to vehicles to identify their exact location. In the system view, which is composed of three different categories (fog node localization, fog node characteristics, and fog management), fog nodes have been located in different locations and have varying capabilities in data processing and data storage, and these capabilities are implemented near the network edge. From a system perspective, having fog management entities gives us the overall view of the fog nodes locations and their different capabilities. Finally, the service view provides different types of services for vehicular networks through having specific policies. Also, they considered RSUs as fog nodes and one server is considered as an SDN controller which is responsible for fog nodes orchestration. They compared this architecture with Fast Traffic Accident Rescue (FTAR) and Traffic Management System (TMS) in terms of performance which showed better performance in terms of rescue time. For example, we can have a decrease in travel time of the ambulance to get to the accident location. Moreover, we have a decrease in time loss that can be gained through differentiating between all travel flow time and free traffic flow time. There are some challenges, such as fault tolerance of the vehicles. For example, when one car goes out of the coverage area of RSU, and it’s considered to provide services to nearby vehicles, the other set of vehicles should assume the responsibility to provide services to the vehicles that still remain in the coverage area. The other challenge is how we can orchestrate fog nodes and how we can manage them in terms of QoE, QoS and etc. Another problem is how we can motivate a vehicle to share their storage and processing capabilities to other cars. In other words, what incentives would work most effectively for us in different policies? Figure \ref{picture1} shows the overall architecture of their idea.

\begin{figure}[htbp]
\includegraphics[width=0.5\textwidth]{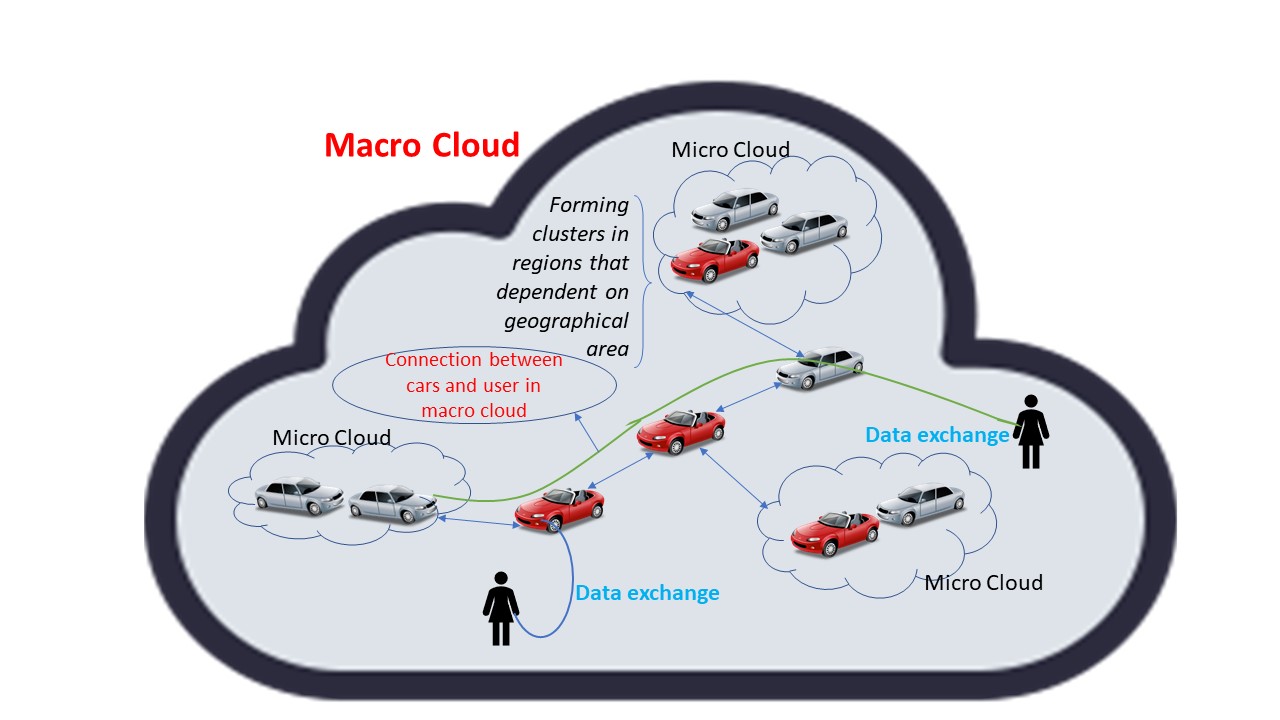}
\caption{Architecture of Macro and Micro cloud in the VANET perspective}
\label{pic-mac}
\end{figure}

	In \cite{luo2011geoquorum} the authors talked about different ideas in macro and micro clouds (MMC). As they mentioned, micro clouds can be formed in some specified regions, such as intersections. Each micro cloud has a management entity named membership protocols. It can be said that the locations of the micro clouds are always the same and the number of members in each micro cloud can be added or decreased. All the cars that are in the same micro cloud are collaborating with each other for providing various services, such as monitoring the traffic status in the regions. Also, If one car left the micro cloud, there are some hand-off mechanisms in place which allow the remaining cars to take over it's responsibilities. In macro clouds, they use the idea of Car4ICT that each car can connect to other cars through exploiting V2V communication to explore the services. This architecture has some advantages, such as: having the static location of each micro cloud give them more power in setting different policies among the cars. Moreover, having a macro cloud allows the cars to find other services that are nearby micro clouds locations. MMC has different phases for storage services. Firstly, it is looking for the best location for installing micro clouds, especially the spots that have enough vehicles, then, it searches for resource availability based on the historical data that has been gathered by servers. Then, macroclouds calculate the amount of stored data in each micro cloud. Finally, they run storage on micro clouds that have to meet the requirements by using a method that is alike to GeoQuorum. Despite some advantages, MMC has some challenges such as incentive mechanisms for cars to participate in the service offering, heterogeneity of the network which is challenging especially when you have to decide which types of communication (such as LTE, DSRC and etc.) and infrastructures (RSU or BS) would work better for satisfying our requirements in different scenarios. Furthermore, incentive mobility can be offered as a solution to the moving cars by keeping connectivity between cars. In other words, how you can motivate the users to choose the path you are giving them to maintain the connectivity between V2V. Figure \ref{pic-mac} depicts the presence of Macro and Micro clouds together in the VANET environment.

\begin{figure}[htbp]
\includegraphics[width=8cm, height=7cm, right]{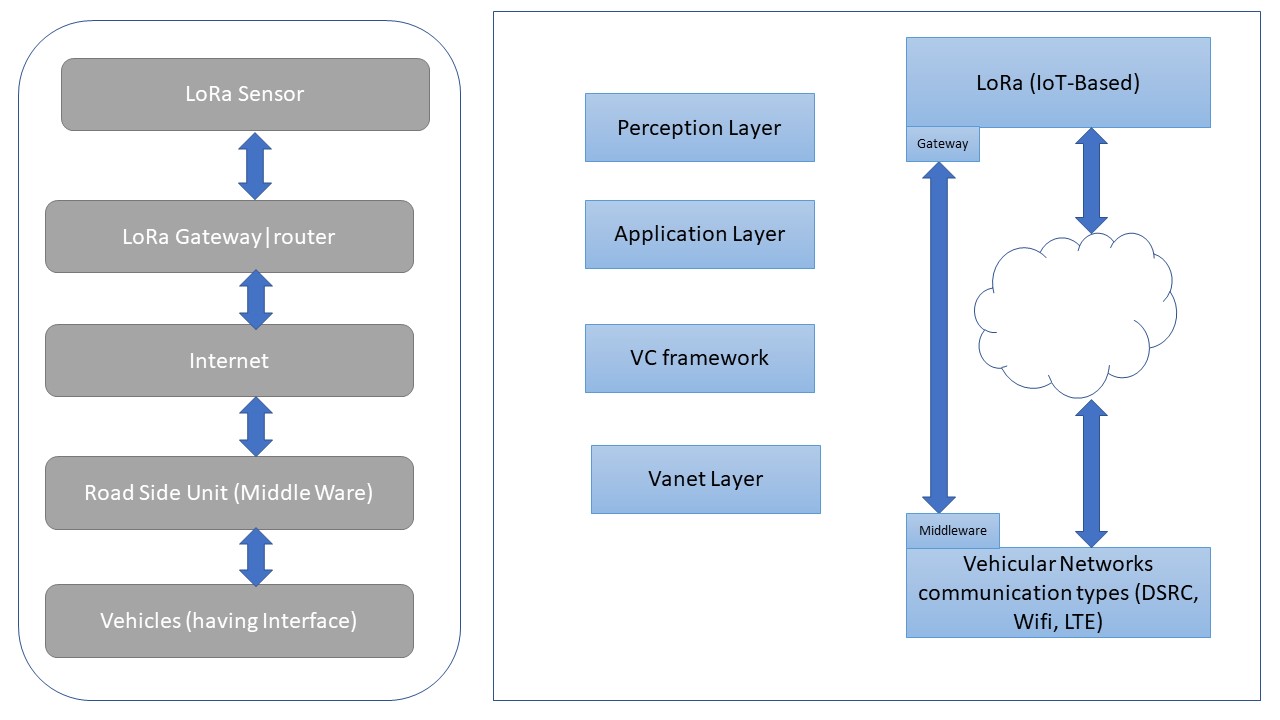}
\caption{Service enhancement through using Lora in a layered design of VANET}
\label{Lora}
\end{figure}

In \cite{khattak2019toward} they proposed the architecture named VCoT (short form of Vehicular networking clouds with IoT). They use LoRaWAN \cite{sornin2015lorawan} as one of the new communication technologies which gives the users long-range communication possibilities. There are many IoT applications that are using different current communication technologies such as 5G, LTE, DSRC, Bluetooth, ZigBee, and LoRaWAN. Its worth noting that, compared to the other technologies such as PRMA \cite{nanda1991performance}, Symphony Link \cite{ramachandran2008symphony} or SigFox \cite{vejlgaard2017coverage}, LoRaWAN has been adopted  with greater success than the others in terms of security, service dependent location, mobility, and communication. They have proposed a vertical architecture in which all the nodes in IoT use LoRaWAN in IoT infrastructures. Also, all nodes connect to RSU to both retrieving data from the cloud and offloading data to RSU as well. As they said LoRaWAN can provide bidirectional communication such as minimal latency, QoS, and better connectivity. They have an architecture which includes three different parts: IoT infrastructure, vehicular cloud and middleware parts. Middleware parts have been recognized through bridges, gateways, and other nodes that have similar capabilities. Cloud parts are responsible for storing and managing information of IoT applications. In their architecture, RSU can play an important role like an interface between LoRaWAN nodes and LORAWAN gateways. In addition, all of the communications are being converted to be considered compatible communications for using LoRaWAN. They showed a use case in their architecture to locate the nearest hospital for emergency vehicles. Also, They used different parameters for each user such as ambulance ID, type of emergency, location of both ambulance and hospitals, and finally criticality level of each injury. RSU, which has enabled LoRaWAN communication, is responsible for getting the live information from patients and the current location of the ambulance, while the cloud component is looking for the nearest hospital for that ambulance. There are some challenges such as extracting useful data from a huge volume of data that are being flooded to the cloud parts. Furthermore, dealing with the heterogeneity of the data would be another challenge, especially when we want to integrate LoRaWAN communication type with VANET communication types, such as DSRC or cellular. Data quality, data coverage, security, and privacy issues are other worth mentioning challenges in this architecture. All in all, figure \ref{Lora} shows a better understanding of their approach.

\begin{figure}[htbp]
\includegraphics[width=0.5\textwidth]{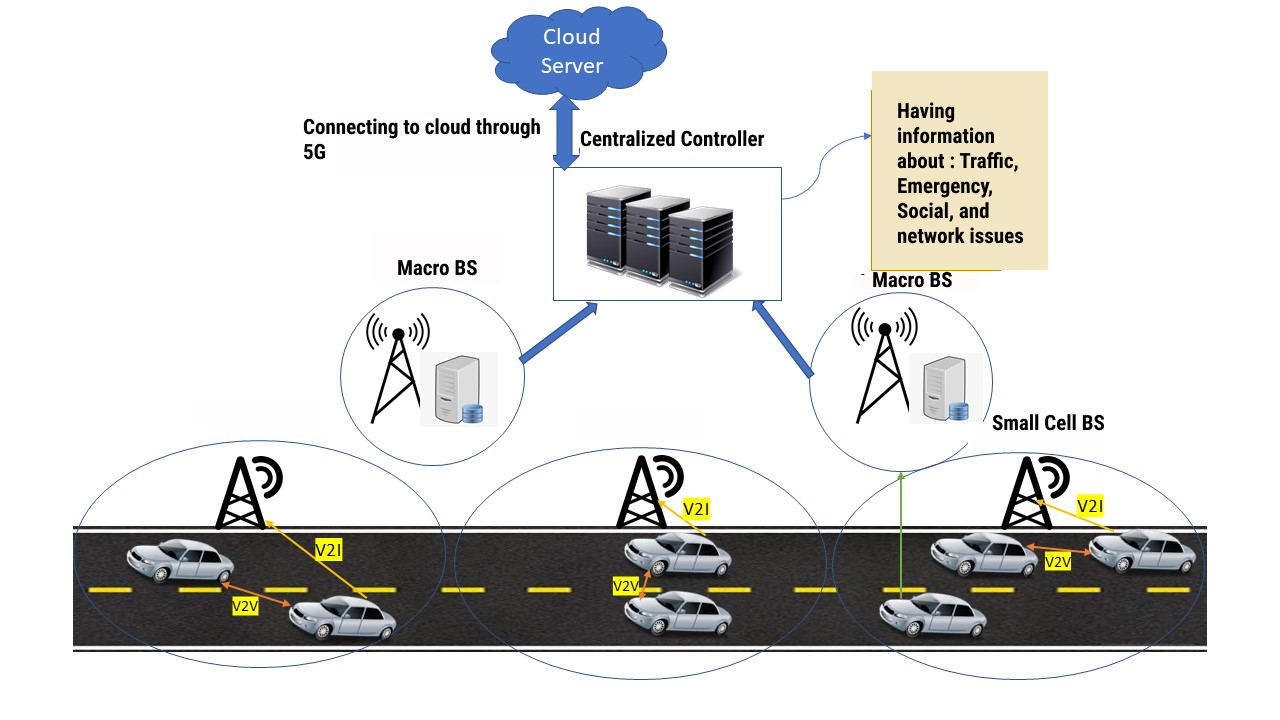}
\caption{Having macro cell BS and small cell BS in a centralized controller}
\label{picture4}
\end{figure}

In \cite{sun2016edgeiot}, they proposed an architecture named edge IoT in which fog nodes can be placed at the edge of the cellular network or they can be connected to BSs directly through fibres and other types of high-speed communications. They used SDN to be more flexible and efficient in managing and storing data. In order to achieve flexibility in the network, they used OpenFlow switches \cite{rotsos2012oflops} which are ruled based on open flow protocols \cite{benton2013openflow}\cite{mcgeer2012safe}. Also, in this architecture, each node can be connected to the cloud by Internet, especially when they do not have enough storage capability, they can connect directly to the cloud to offload their data on the cloud which leads to higher cost of resource usages such as Internet access, using cloud storage, consuming network bandwidth etc. We can have some IoT applications which are deployed in both fog nodes and clouds to provide services to the users. In this architecture, each user can have a proxy VM that can be placed in the nearest fog nodes that are responsible for data computing and storage. In this sense, all the devices that belong to the users are being registered into their proxy VMs to collect data from IoT user devices to group them based on the data type. After analyzing the data, the data are transferred into metadata and provide worthwhile information while maintaining user privacy. For example, when VM proxy receives an image, it only takes the location and the time that the photo was taken instead of saving the original photo into its database. Furthermore, this picture can be received from different VMs which are owned by different users. Also,  we can extract the metadata through analyzing raw data that are coming from different proxy VMs and offer useful services to the users. As each user can have IoT devices which are either dynamic, such as mobile phones, or static, such as electronic thermometers in their homes, VM Proxy can be decomposed into two proxy VMs, one of which serves the static IoT devices and the other moves along the dynamic IoT device. They concluded that the location of each VM based on our requirement can be different. They can be deployed either locally for analyzing specific locations (for example finding parking spots in a certain location) or remotely in the cloud (for analyzing the larger area of the city). In some cases, when we have a triggered event, such as terrorist attack or child missing, we can deploy a VM application in the requested area to analyze the historical data in that location. There are some drawbacks according to this architecture. First, whenever the proxy VM wants to perform composition/decomposition, it has to inform all registered IoT devices. Moreover, the users have to know about their IDs whenever they want to send data to proxy VM to make sure it is an accurate proxy VM. Furthermore, increasing the traffic load of the network through migration, especially when we have fast mobility of the user devices, can be considered another challenge. For instance, if a user wants to leave BS1 (Base Station 1) and goes to BS2, obviously her proxy VM moves along with her to get raw data from the device. If this moving takes “t” time and before migration completion, the user moves out of BS2 coverage, this is an ineffective migration completion. It interferes with controlling bandwidth and traffic of the network inefficiently, as it has produced extra traffic for the network.
Luo et al.(2018) in \cite{luo2018cooperative} proposed an architecture that works in a distributed manner in terms of storage and controlling entities. All the nodes throughout the network are cooperating with each other to share raw data and make improvements in data sharing. There are two layers, namely the upper layer and lower layer. The lower layer is named control and management layer or macro base station (MBS). This layer controls and manages a smaller area which includes RSU, Vehicles and Base Stations (BSs), or Wi-Fi nodes. Moreover, MBS supports efficient cooperation among different types of communication such as cellular, DSRC, and Wi-Fi. For example, cellular links can be used for transmitting control information and Wi-Fi, and DSRC can be exploited for content transmission. MBS can be considered the best place for running heavy algorithms and then release the best decision based on the output of the algorithms and at the same time vehicles and RSU can offer services to the requests of cars. The upper layer named CWC (short for City Wide Controller) is responsible to retrieve the data that are coming from different MBS. For example, each MBS can have local traffic data and CWC can easily retrieve all of these data from MBSs and aggregate them. Based on historical data and real-time aggregated data from MBSs inside CWC, CWC can predict the traffic in different locations. There are some challenges such as the rapid topology movement of the network and the large volume of data that needs to be mined quickly in order to extract useful data. Moreover, many vehicles request a large volume of data, such as watching a movie or playing a game or etc. One of the solutions is prefetching data in the BS and RSU in order that the cars get access to them as fast as possible. In this way, we need to have an efficient and flexible strategy to satisfy high speed car requests to replace their connection with BSs and RSUs frequently in unbalanced traffic on the roads. Idle time is taken into consideration as well. Regarding the idle time, we can have a multi-factor- multi-place strategy to prefetch the required data by cars in a suitable BS or RSU. For example, the idle time of cars in the gas station or high traffic spots is more than in other locations, so they can connect more time to the same infrastructure in those regions. Furthermore, prefetching data, might have a better result in terms of minimizing latency and decreasing the high data load of the network. When the data is prefetched in a desired location, MBS schedules different types of communication modes to have cooperation with each other. This issue gives rise to complexity in cooperation terms, especially when the channel capacity is low and full. Moreover, they have said using cars as resources for prefetching data can be another solution. All of these possible solutions can be gained through machine learning algorithms to predict which places might have high potential traffic. Their idea can be partly summarized in figure \ref{picture4}.

\begin{figure}[htbp]
\includegraphics[width=0.5\textwidth]{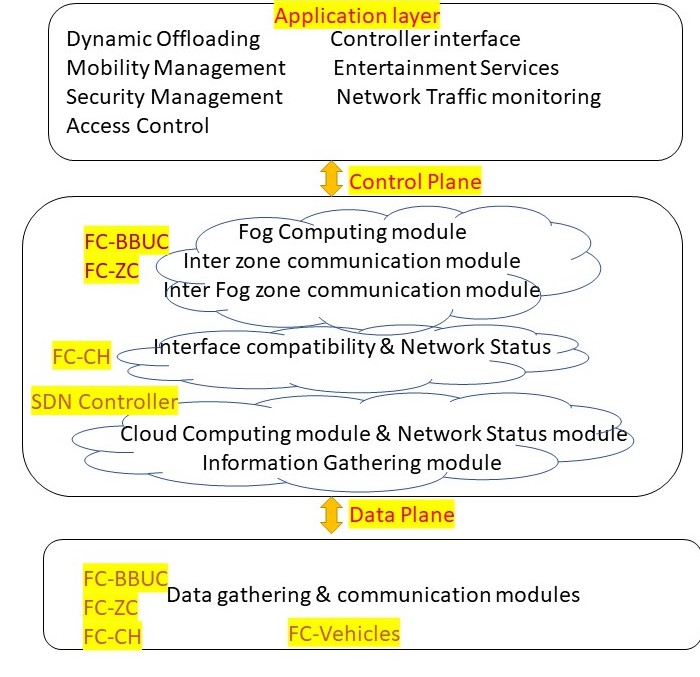}
\caption{placement of FC-BBUCs, FC-CHs, FC-ZCs and SDN controller in the architecture}
\label{picture5}
\end{figure}

Khan et al(2018) in \cite{khan20185g}, proposed a combination of fog, SDN, and C-RAN. Fog computing is composed of different entities, namely Fog Computing-Zone Controllers (FC-ZCs), in which a zone has a number of vehicles which have been registered by BSs or RSUs in that region. Each zone is being controlled by SDN controller. It's important to know that in this architecture, all the controlling data are not being sent to SDN controller, and some controlling overhead of cars are being kept in their own vicinity. The other term is named FC-CHs, which describe the cars who have their OBU which has SDN. Fog Computing Base Band Unit Controller (FC-BBUC) is another term which is responsible for connecting multiple FC-ZCs through backhaul links. There are some advantages to exploit CRAN in this architecture, such as remotely controlling resource allocation and managing scheduling globally in a simple way. In other words, it can be said that FC-BBUC can be considered as a bridge which connects SDN controller to VANET. It has its own management entity for decision making as well, and in many cases, it doesn’t need to send the data to the centralized entity for making decisions locally. If each FC-ZC wants to get the intercommunication inside of the zone, it first arranges it with FC-BBUC. It can be said FC-BBUC plays the role of both control and data plane devices. Another term is SDN controller which controls the network metrics, such as making a decision about resource allocation, mobility management, and so on. In this case, SDN controller is distributed throughout the whole network in a hierarchical way. Another responsibility of SDN controller is Fog orchestration. In addition, another term is named Optical Transmission Network (OTN) which provides the link between FC-ZCs and FC-BBUCs. So, when each FC-ZC wants to connect directly to FC-BBUC, optical fiber comes into play and digital signals are being sent through it. This architecture has three different layers: data plane, control plane, and application plane in which FC-Vehicles, FCZCs, FC-CHs, and FC-BBUCs are put into a data plane. They aggregate data through the information gathering module and then send them to control plane through the communication module. The information-gathering module collects all the data that are coming from sensors on the cars and so on. The communication module provides different types of communications such as V2V, V2I and connection between FC-BBUC to FC-ZC. As there are fog architectures at the edge, and near our devices, SDN controller will be run in a hybrid manner. In figure \ref{picture5}, the control plane consists of FC-BBUCs, FC-CHs, FC-ZCs, and SDN controllers. SDN controller sends some rules to each FC-BBUC and FCZCs in an abstract way or general rule, and these entities can make the specific rules based on the policies that they have in their own region. They compare their throughput, transmission, control overhead and some other factors with traditional architectures, in which each car connects to the closest RSU to send data. The charts showed improved performance in all the metrics. They have not mentioned the challenges they might consider for their architecture, but as it is clear, analyzing big data and making decisions rapidly for emergency messages are considered vital. In addition, heterogeneity of the network and hands-off between vehicles and RSU or BS are other challenges that have not been addressed in this paper.

\begin{figure}[htbp]
\includegraphics[width=9cm, height=7cm, right]{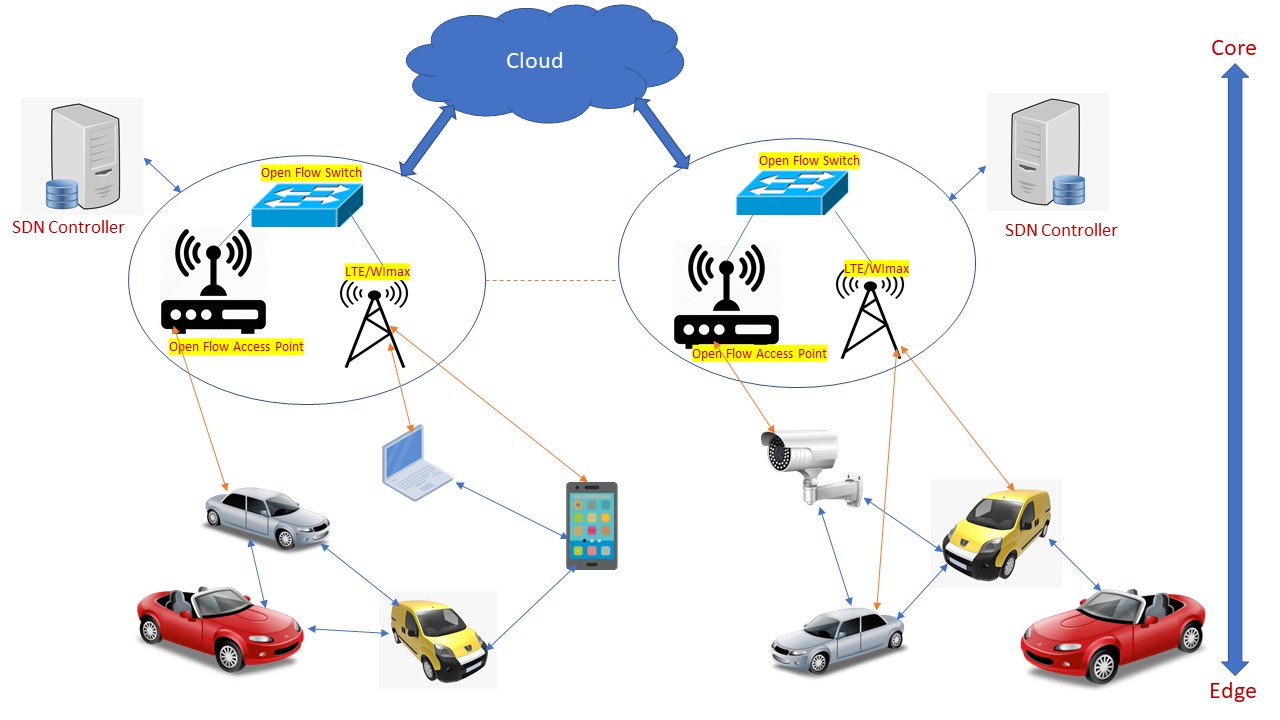}
\caption{Having distributed OpenFlow switches and SDN controller coupled with cloud}
\label{picture6}
\end{figure}

Tomovic et al.(2017) in \cite{tomovic2017software} showed a structure that is composed of SDN controllers, fog nodes and cloud infrastructure. For managing resources and deployment of IoT applications, fog nodes can have a set of APIs for accessing a number of different resources, such as different operating systems in virtual machines. Moreover, they manage other physical resources remotely. The mobile fog computing model can be used to simplify the use of different types of IoT applications. Through deploying fogs in a mobile manner, we can run the same IoT application in different fog nodes, in which each node is responsible for a number of tasks. For example, surveillance detection application can be done in three steps: movement exploration in security camera, face detection at fog nodes and data collection and analysis which are run in cloud server. All these fog nodes know about their specific tasks regardless of their knowledge of running the same application and their locations. One of the challenges in the mobile fog model is service orchestration, which includes service migration or service instantiation on a huge number of fog nodes that have a variable number of capabilities. As there are many event triggered applications that need to immediately allocate resources and data transferring in run time, a centralized SDN controller is essential for fog orchestration. SDN controller is responsible for some tasks, such as fog orchestration and finding the best RSU or BS for IoT devices to be connected to. SDN controller needs to update its data frequently to get current fog status, such as how much RAM storage is available or what operating system each fog node is using currently. Moreover, SDN controller can find out how congested each link is, how much data has been loaded on each link, and so forth. In terms of business perspective, each fog node can have the minimum requirements of each application that are offered by businesses. Each company specifies the resources, storage and computation power needed to run the application correctly and fog nodes are responsible for providing this level of requirement. In order to provide required resources and application policies, SDN controller needs to know about them, especially when it wants to allocate resources and manage the resources for allocating them to fog nodes. Based on some predictive algorithms, SDN controller can predict the location of moving vehicles in the near future and allocate enough resources to the regions that have more potential to experience traffic. Another role of SDN controller is controlling the traffic by getting real-time data from different fog nodes. It's worth mentioning that SDN controller exploits OpenFlow controller for traffic management. As each SDN controller has been installed in the upper level of fog nodes, it can manage multiple fog nodes that are placed at a lower level. There are some local SDN controllers in fog nodes to provide robustness. If our application serves emergency services, we can split the data and data management and analyze data with different priorities. The fog nodes send their status and device information to the SDN controller. If SDN controller comes to the conclusion of having unbalanced traffic in some areas, it recalculates the route proactively to decrease the pressure of unbalanced traffic in processing nodes. They \cite{tomovic2017software} showed some use cases, such as smart transportation and video surveillance. In this structure, a huge amount of data is being processed and analyzed in SDN controller and fog nodes send a high load of data to SDN controller for further analysis. It can be said that in cases in which delay plays an important role, this architecture cannot be effective even if we prioritize the IoT applications based on their needs for accessing resources.In figure \ref{picture6}, they clearly show the distribution of devices from edge to cloud.

\begin{figure}[htbp]
\includegraphics[width=0.5\textwidth]{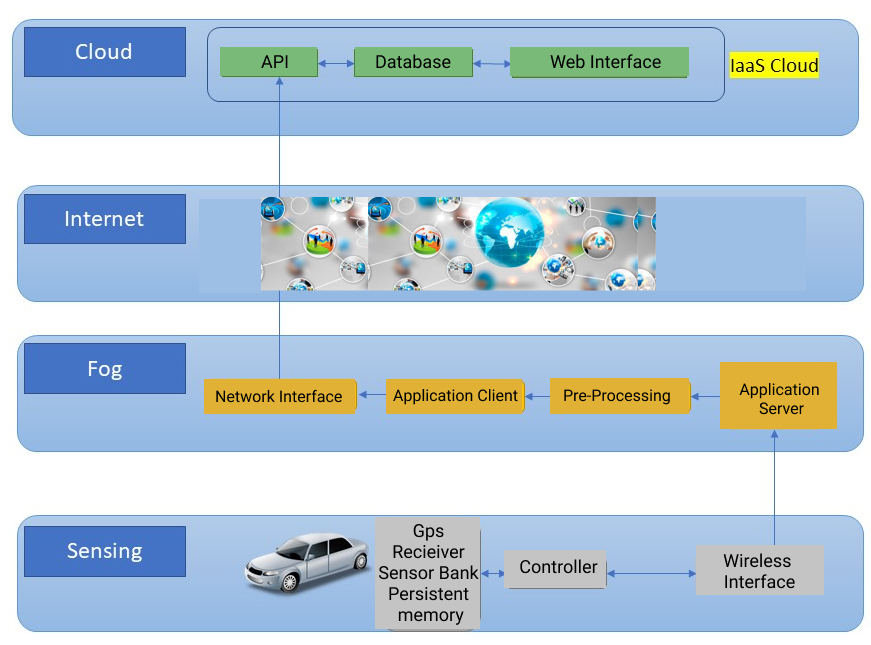}
\caption{3 layer architectures in sensing Bus}
\label{picture7}
\end{figure}

Cruz et al.(2018) in \cite{cruz2018sensingbus} proposed an architecture named sensing bus which has some main tasks namely: data gathering, pre-processing of data, data transmission, data storage, and data service. In data gathering, the sensing bus collects the data from each bus. Then, data is being pre-processed before sending it to the network to make sure channel bandwidth is being used with useful data, not malicious data. In data transmission, buses can send the data to the users and users evaluate the correctness of data. In data storage, all the data goes into a database which is placed into the cloud. Finally, in data service, the sensing bus offers different services to the users. Moreover, there are 3 different layers in the sensing bus, namely: sensing layer, fog layer, and cloud layer. The sensing layer should be placed in locations where there is frequent vehicle activity, such as bus stops. They sense the environment and gather data, then send the data to the fog layer. Sensing nodes also save the time stamp and geographical location of the recorded data. Fog nodes are installed in different locations of the city and are waiting to receive data from the sensing layer. Their locations are fixed on the edge of the network, such as bus stops. Once the data are sent by sensing node to the fog level, all the data in sensing nodes will be deleted. The fog layer, pre-process the raw data which has been sent from the sensing level and then send the data over the internet to the cloud. Fog nodes can operate in both client and server roles, depending on which layer we are considering. For example, sensing nodes see them as access points and also see them as application servers, which retrieve the data from them. On the other hand, the nodes in the cloud layer see fog nodes as application clients. Communication between fog and sensing nodes can be done through a private network. Moreover, the connection between fog nodes and cloud level can be done over the Internet. Inside the cloud, the data is being pre-processed, analyzed, and finally made available for the users. As cloud level has more computation power for running heavy algorithms, the best place for analyzing the received data is in the cloud. Another reason is availability, for example, if we want to broadcast the warning messages to the network, we can do it from Cloud layer. There are some good points in this paper, such as working with real data in Rio de Janeiro in Brazil, which showed that each fog node can serve up to 20 sensing nodes without dealing with computation and memory shortage. It's important to say that, if the number of sensing nodes increases, the throughput in each of them decreases because of competition for channel access. However, this architecture only works when the number of sensing nodes is about 20 or slightly higher. So, dealing with big data through this architecture without having orchestration between fog nodes and without taking consideration of unbalanced traffic in some areas is extremely hard. In figure \ref{picture7}, they showed their architecture with having three layers.

\begin{figure}[htbp]
\includegraphics[width=0.5\textwidth]{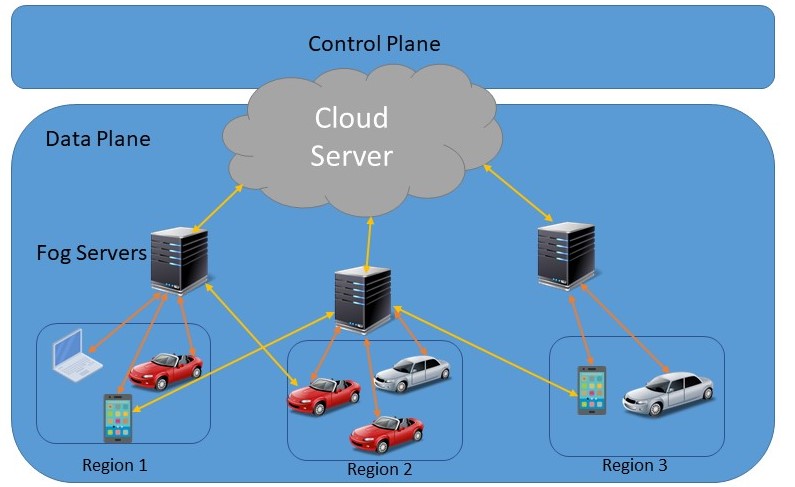}
\caption{Having a hybrid model in Fog computing}
\label{picture8}
\end{figure}

Ning et al.(2018) in \cite{ning2018mobile} discuss a model that exploits NOMA (non-orthogonal multiple access)\cite{saito2013non} and MEC (mobile edge computing). There are some major components in the model such as: RSU, vehicles, and macrocells. When one event is triggered, it is being recorded by the vehicles and users, in different data types. After that, all data is aggregated by the cars before offloading. There are cluster heads between the group of vehicles and each of them is responsible for collecting data from its members. Also, there are different costs for message generation by different entities. For instance, if the message is produced by macrocells, there is a little delay that can be ignored, but there is a high cost which is coming from the mobile operator. On the other hand, if the message is produced by RSU, there is no cost, but we have a delay in data transmission. In this network, one subchannel in NOMA can be achievable by different vehicular users and one packet can be received through different subchannels from the vehicular users. Moreover, to cancel interference, transmit power control has been proposed, in which, a user with better coverage of microcell, has to send their data through the lowest transmission power level. Also, when they are not in a macrocell coverage area, they send their data with the highest transmission power to guarantee data delivery. Moreover, vehicles can offload some of their heavy tasks that need more computation to other macrocells or RSU, however, offloading messages to other entities causes additional costs for the network in terms of bandwidth consumption and delay. For example, it takes time for the messages to be offloaded to the RSU and returned to the vehicular user after data analysis. This issue is more critical in the delay-sensitive applications. For calculating the delay, the author showed some policies to choose either RSU or macrocells for data offloading by the users. If the vehicle arrival rate and their travel time are bigger than the threshold, vehicular users choose macrocells for their data offloading, otherwise, they prefer RSU for data offloading. In terms of power allocation and assigning a suitable subchannel to each user, a hybrid scheduling model is proposed, in which, vehicular users adjust their transmit power in a distributed manner while macrocells do the scheduling in a centralized manner. When macrocell gets the real position of each vehicular user, it can specify the resources of frequencies for each one. Furthermore, we have co-channel interference in NOMA as well. After evaluating their performance in the real world example, taxi routes in Shanghai, China (2015), they concluded good performance in traffic offloading and transmission rate gain through having vehicle speed, vehicle location through GPS, and having record time. They compared their model with random offloading and F-RAN (Fog Radio Access Network). Although this paper did not show any fog and cloud architecture, it takes a traditional architecture of vehicular networks through having edge devices as fog nodes. Moreover, each macrocell can behave like RSU. One disadvantage is that the macrocell status, when it's overused in terms of storage and computation, is not taken into consideration. Another disadvantage is interference between other macrocells, which has not been considered in this paper.
\newline
In \cite{gao2017fog}, the authors proposed a simple model in which fog nodes cooperate with each other. In order to offload data between fog servers, they used DTN (short form of delay-tolerant network). If one fog node has small data and the other fog node doesn’t have that content, they can directly connect to each other and share or offload their data through using DTN technique, without connecting to the cloud. For instance, consider a user who is sitting in one store and download his favourite movie from fog node A. After a couple of hours, he decides to go to other location which is in the fog node B coverage. This user can move the downloaded data from fog node A to fog node B through his mobile phone that already has the downloaded data, and the data can be offloaded to fog node B through his mobile phone. As this is obvious, through the process of store-carry-forward, fog nodes don’t need to connect to the cloud to get the desired data that exists in one fog node. For offloading a large volume of data that is not possible to be stored on the mobile phone, parked cars can be used. Apart from data transmission with DTN technique, there are direct communications between fog nodes as well. For example, a bus can have a fog server, and passengers on the bus can easily get fog server content through a wireless connection. In some cases that data are unavailable for the users, fog server immediately connect to the internet to get the data for the users. If this bus moves along the roads and finds other fog nodes, first it synchronizes its data with them and if the fog server on the bus doesn’t have some data, it downloads the related data and then check other fog nodes along its way to make sure all of them have been synchronized with each other and have the same data as well. Fog servers and cloud servers cooperate with each other and if some data are unavailable in fog nodes, Cloud server provides the required data for them. In addition, Cloud server acts as a control plane to evaluate whether data is needed to be uploaded in fog nodes or not. All the fog nodes and a part of cloud servers have data plane for data dissemination. Data dissemination has three parts, namely, Data structures, algorithms, and protocol messages. Data structures are responsible for finding the path for data transmission. Protocol messages are used to explore content and the devices belong to each fog server. Finally, algorithms calculate the best link for data dissemination. Figure \ref{picture8}, depicts their hybrid model.

\begin{figure}[htbp]
\includegraphics[width=0.5\textwidth]{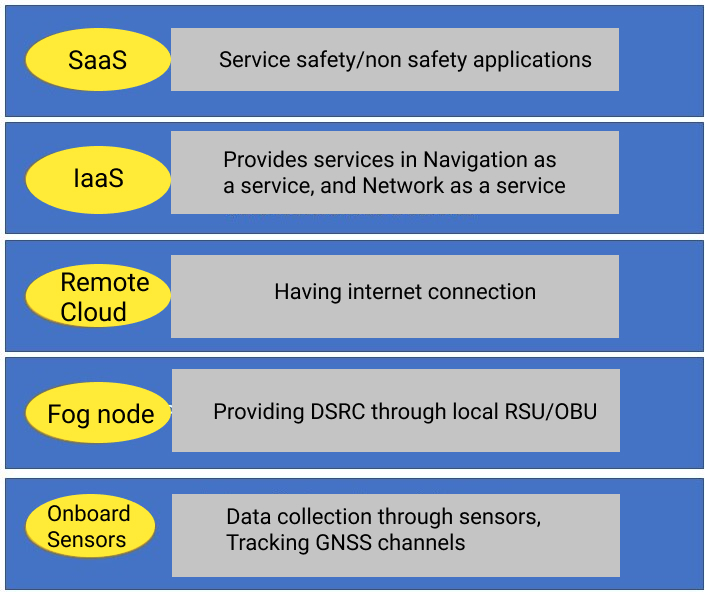}
\caption{C4S architecture layers }
\label{picture12}
\end{figure}

In \cite{shah2019vfog}, they proposed the architecture that is focusing on V2V communication. There are some vehicles that share their computing and storage capabilities, named Vfog, and other cars can request to access their capabilities. After that the requests are being received by Vfogs, they are being served by 'first come first served' architecture. For responding back, Vfogs have the capability to keep track of requests address. If their current locations are completely out of Vfog coverage area, the ad-hoc model is exploited to send the messages. There are some limitations to each vehicular network. One of the important ones is the fading effect, which clearly can reduce the entire performance of the model. In order to decrease this fading effect, Nakagami distribution model \cite{yacoub1999higher} has been used in this architecture. In a practical case, they showed that fog vehicles frequently inform the other vehicles about their capacities. Multiple options for vehicles give them this opportunity to choose the best Vfog, which can satisfy their requirements, such as fast delivery and desirable connection. Before sending the requests, in order to set up the path between them, a handshaking mechanism is established between cars and Vfogs. After the handshaking process, vehicles are informed about queue length and expected response time.  In this model, when the number of vehicles is increased till 120, the average transmission delay and packet delivery ratio are increased as well and obviously lead to a congested network which interferes with a quick response time. 
\newline
In \cite{lai2018fog}, lai et al. assume that each car monitors the road status and surrounding area with periodic sensing. Through different cycles, they send all the information to the RSU through direct communication. Each RSU can be considered as a fog node in which, high computation and storage power are available. Moreover, each RSU can trigger an event detection to test the received data. If they need more data for more analysis, they get the data from cars or Base Station (BS). All these data are going to be uploaded into a central server which has more knowledge about the road status and finally makes a decision based on the data analysis. Also, there are 3 different jobs for the event-based data collection model: firstly, when an event is activated, the central server is informed about the event through gathering data in the environment by sensing. Secondly, after sensing some close occurrence events, data gathering is started which has to meet time limitation and delay issues. Finally, the central server evaluates the event existence and provide information to our network. In terms of sensing, there are two types of sensing modes: low cost and high-cost sensing or deep sensing. The first one needs low computation and storage power and the second one needs more computation and storage capabilities. For example, when a car is in a location that has been deformed on the pavement surface, it predicts the event with 50 percent confidence. But, in order to confirm that this event has a high probability to happen, other cars that pass through this surface should declare their prediction for that event. Finally, based on the prediction of all the cars, and more analysis of that event inside the RSU (which is considered fog node as well), the final decision is released about the chance of having that event. In their model, they supposed that, sensing can be done by taking pictures of the roads every 30 seconds and the quality of the picture varies based on whether we are sensing the road deeply or not. There are some challenges in this architecture as well, such as not having incentive mechanisms for sensing tasks by cars and dealing with the channel that would be full when the cars send their sensed data to the RSU at the same time.
\newline
Ansari et al.(2018) in \cite{ansari2018cloud} proposed a method named Car4S which has 2 layers of Vehicular Cloud Computing (VCC) model. There are local fog nodes and remote cloud. Moreover, they considered onboard infrastructures and IaaS/SaaS layers.Also, limited and different capabilities of storage, and computation of onboard units is taken into consideration. So, when they need more capacity for analyzing, they send their request to the fog or cloud. Each fog node provides data access near the users to minimize the delay as much as it can. The bottom layer of the proposed model is onboard units on the cars, with three different types of communications namely: DSRC, vehicle to everything V2X, and cellular. In the third layer, Information as a service (IaaS) is placed to give the users real-time data. Also in Network as a Service (NaaS), it provides information about fog node local availability to users. Furthermore, the cloud layer is available in this architecture, and in most cases that the messages are not time-dependent, the analysis is performed in the remote cloud. In their navigation as a service, they offer two new terms namely manoeuvre planning and execution which helps the drivers to change their lanes safely. NAVaaS (Navigation as a Service) is run in two main key components: the first one is the Onboard Unit (OBU) which collects data and sends them to Manoeuvre Application (MA). MA has been implemented on local fog or cloud nodes. After getting data from OBU by MA application, as it has access to GNSS (Single and Multi global Navigation Satellite System), it can correct some parts of the navigating system or even offer better estimation in location and speed of each vehicle. They showed the navigation in both DSRC communication and Cellular communication with a different velocity of the vehicles and finally concluded that with increasing cars speeds, transmission delay will be increased as well. Figure \ref{picture12} demonstrates the five proposed layers of C4S.

\begin{figure}[htbp]
\includegraphics[width=0.5\textwidth]{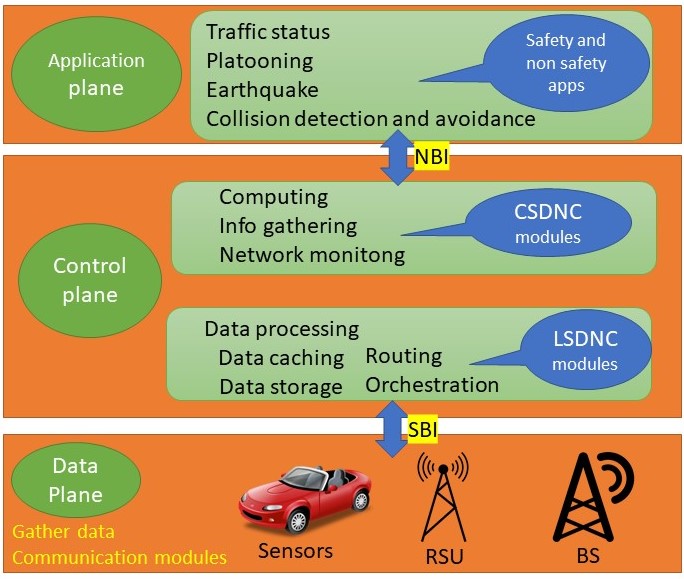}
\caption{ Multi layer of 5g fog SDN based in VANET }
\label{picture10}
\end{figure}

Xiao et al.(2019) in \cite{xiao2019fog} represent a hierarchical architecture in the VANET network. They proposed that each RSU can be considered as fog nodes when they have a smaller coverage area. Moreover, the RSU and other infrastructures that have larger coverage areas are considered as cloud nodes. Cloud nodes are directly connected to SDN controller through backhaul links. Intra fog cooperation and inter fog cooperation are two different terms for sending the data between different layers. In intra fog cooperation, each fog node sends the data to the users that are inside its coverage area, through V2F (Vehicle to fog) communication, and each user inside the coverage area of that node, can send the cached data to other vehicles that are in the close vicinity of itself through direct communication (V2V). In inter fog cooperation, fog nodes share their data through the scheduling mechanism of the cloud. They evaluated their performance through Average Service Delay, Service Ratio, and Bandwidth Efficiency. When the number of requested data packets growing, average service delay is growing as well. However, for cooperative services, average service delay increases slightly. Compared to NBC and MA \cite{liu2017coding}, cooperative service can serve more requested data when the workload is high, which means that this system has a high service ratio. This architecture has some challenges as well. Although it supports scheduling mechanism in the cloud, it does not show about the emergency messages that need to be analyzed as fast as possible, In other words, they do not have any message prioritization.
\newline
Paranjothi in \cite{paranjothi2018dfcv}, shows a three-layer architectures namely terminal layer, fog layer, and cloud layer. This model supports one to one, many to many, and one to many connections. This architecture mainly focuses on data transmission through different scenarios, namely merge and split. In the split scenario, through a function named distance, the distance between cars is achieved and then the entire capacity of the model is calculated through different functions. If the distance and capacity of the model is greater than the threshold from the sender view, the split mechanism for relaying data is  used, otherwise, merge mechanism is chosen for data dissemination. They showed their performance in urban and highway scenarios. In the highway scenario, the chance of splitting and merging take place more frequently than urban scenarios, because distance between vehicles on the highway varies a lot compared to the urban environment. Apart from having good management in resource allocation and communication orchestration between end-users through the split and merge approaches, this architecture did not consider the time for split and merging mechanisms.
\newline
In \cite{soua2018multi}, Soua et al.(2018) present an architecture through having sub-models. central SDN controller is available in the cloud. In addition, there are some temporary fogs and each of them has one local SDN controller for decision making locally. This hybrid architecture has some advantages such as: analyzing real-time data that are coming into the fog nodes and making decisions by each sub-SDN in an appropriate time. Moreover, there is no need that all the decisions are being made by the central SDN controller which resides in the cloud. clearly, it enlightens the load of the data that are going into backhaul links as well. Moreover, all the local SDN controllers are in the control of the central SDN controller. There are some other terms in this model, such as CSDNC which controls the network, and different behaviours that might happen through different scenarios. Moreover, it manages the resource allocation between fog nodes. The other term is fog cell head which plays a very important role in the fog nodes which has the mobility nature. Creating fog cells and installing SDN controllers are some jobs of fog cell head. Fog cell head is under the control of CSDNC, and they can be run on vehicles such as buses and taxis. Another term is named fog nodes that can form the fog cell, and again it is under the control of LSDNC. Each fog node is equipped with an Onboard Unit (OBU) which support both DSRC and Cellular communications. Finally, base stations are the middle entity between cloud and fog nodes to make sure the communications between them. In figure \ref{picture10}, we can see the overall layers that they proposed.

\begin{figure}[htbp]
\includegraphics[width=0.5\textwidth]{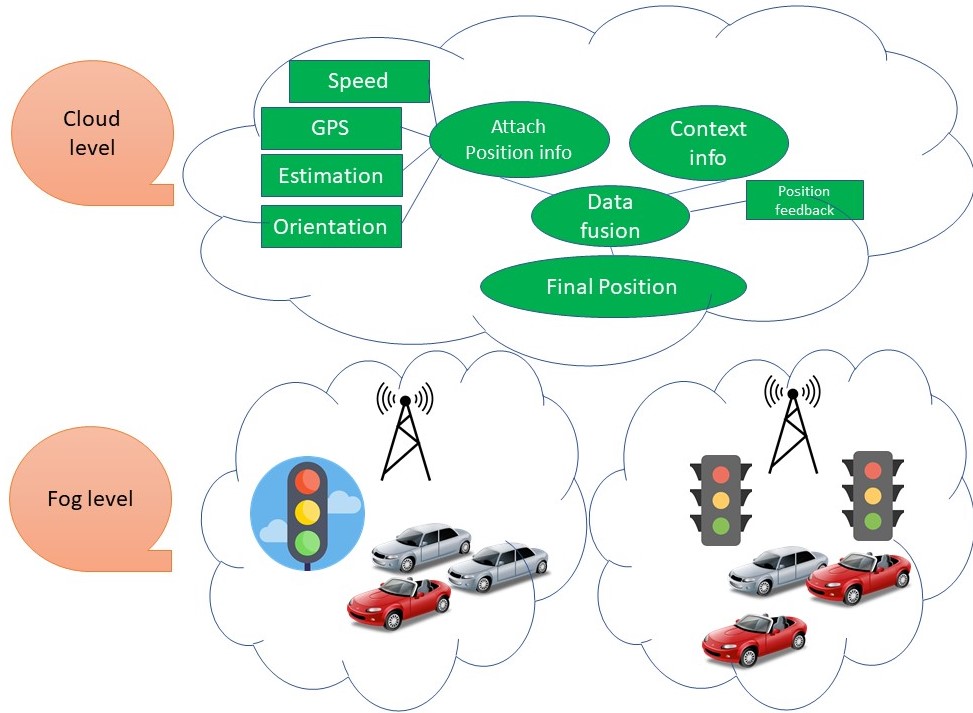}
\caption{ Demonstration of fog level and car localization in the cloud level }
\label{picture15}
\end{figure}

In \cite{khattak2019integrating}, Khattak et al.(2019) talked about VFN (Vehicular Fog Network) which is composed of different layers namely: Vehicular Network, Vehicular Fog, and Service Provisioning Layer. Vehicular network layer includes all the vehicles that are equipped with OBUs. OBUs are responsible for collecting data from the surrounding environment and analyze them in low level and finally send them to the nearest edge node. Vehicular fog layer are the nodes with huge computation and storage capabilities and are installed in some important and crowded locations such as junctions and parking areas. All these nodes have a direct communication link with cloud and vehicles. Finally, Service Provisioning Layer provides different services for the vehicles, as it has enormous computation and storage capacity. It distributes the tasks between fog nodes efficiently.
\newline
In \cite{lobo2017solve},Lobo et al.(2017), gives the user exact location of each car by combining fog, location of the traffic lights and RSU. Their model has some components such as : Vehicles equipped with OBU which already know about their location through having GPS and WAVE (Wireless Access in Vehicular Environment). All of these traffic lights are equipped with sensors in order to find cars locations and send related data to Road side Units (RSUs). Another important component is RSU, which has enough capacity and through using Data Fusion model, it would be able to get the node location and send related information to the cloud. Fog in this model includes RSU and smart traffic lights. Furthermore, in some cases fogs can be composed of not only RSU and traffic lights, but vehicles as well, to provide more computation and storage power. The final term is cloud, which knows about the location of each node and overall view of the network. There are some challenges in fog and cloud layers in this architecture. For instance, in fog layer, as the author mentioned, finding vehicle location with Data Fusion with higher accuracy is hard. In cloud layer, as it has more computation and storage capabilities, it can offer some patterns about the traffic route and offers which paths have low and which ones have high traffic, and direct drivers to the suitable path. In figure \ref{picture15}, it can be clearly seen how data fusion and final position of the cars can be happened in Cloud level.

Brennand et al.(2017) in \cite{brennand2017novel} proposed a system named FOREVER, and it uses RSUs and sensors in a distributed way. First, the city is divided into different areas and each area is equipped with RSU and one fog node. Also, Area of Knowledge (AoK) is a term in this model which has a virtual map. In this section, RSU can provide information about the current status and future routes for cars. In terms of coverage area, each AoK has the same coverage area with RSU and the size of the AoK has a direct effect in computation time prediction accuracy of the a new route in the future. For gathering information, each RSU sends related information about its Aok, the size, and its location to the vehicles to giving them an overall view of what RSUs are nearest to them for connecting. On the other hand, the vehicles send some data such as their location and speed to the RSUs. RSU uses this data to get the knowledge about the behaviour of each vehicle. When RSU received the messages from different Vehicles, first, it checks about how much is related to the AoK, or is this message related to its current AoK or not. If it's related, RSU updates the AoK databases, and if not, RSU drops that message. 
Lian in \cite{liang2019extremely} categorized the nodes based on their mobility and stationary status in 4 different layers.  For example, in the first and second layers, static nodes can communicate with static nodes, in the third layer dynamic nodes cooperate with static nodes in the second layer, and finally, dynamic nodes in the fourth layer communicate with dynamic nodes in the third layer in their architecture. They formulated the problem based on not uniformity distribution of the nodes along the network, limited movement of each node (for example each car only moves along the road not out of the road), and inconsistency of communication coverage and noes clock rates. Most of the nodes in VANET have mobility, and if the node that is responsible for sending the synchronization packet doesn’t send it at the right time, the receiver which has the mobility behaviour, change their location before getting the packets at a right time. So, time synchronization is vital in Fog-based Networks, which has heterogeneous behaviour. They gained a method for precise time synchronization in dividing the node's behaviour into static and dynamic nodes, which doesn’t change significantly if, the network size and clock error frequency increase.

\subsection{Recent works in the context of VSN} 
\begin{figure}[htbp]
\centering
\includegraphics[width=0.5\textwidth]{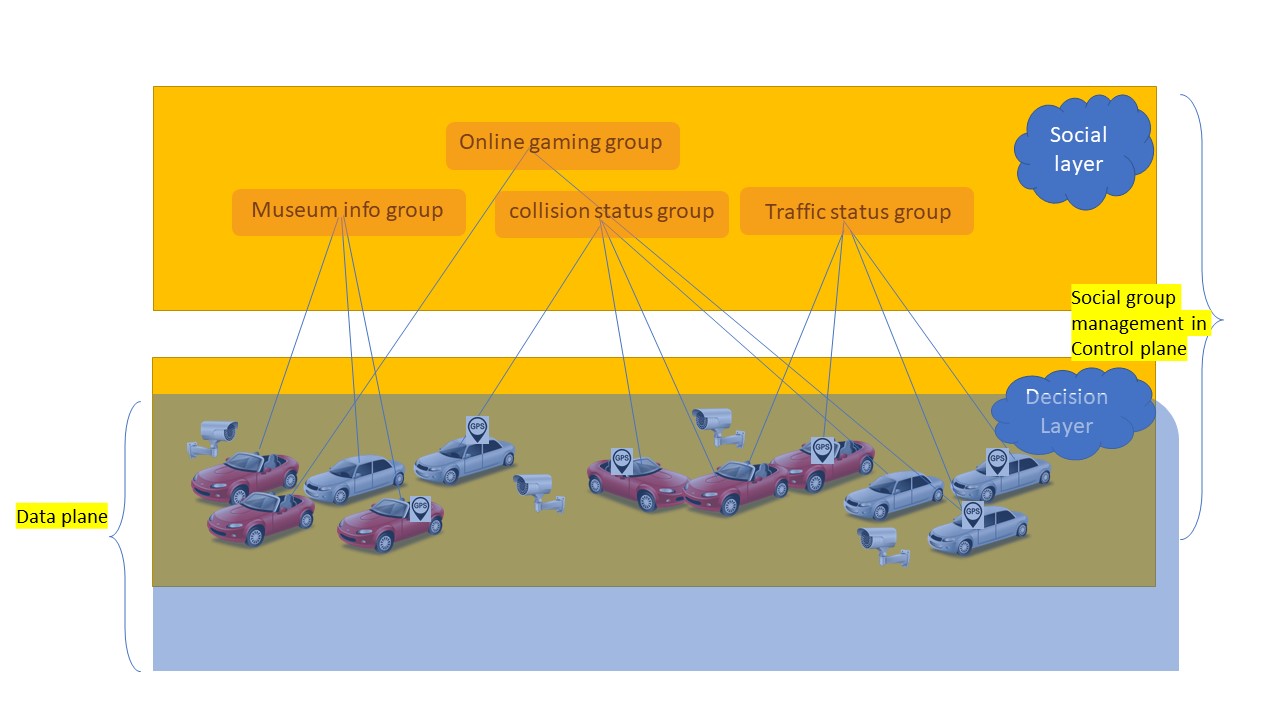}
\caption{ Demonstration of Social layer and Decision layer in the control plane }
\label{picture16}
\end{figure}

In \cite{li2016veshare}, Lie et al.(2016) use SDN to control the messages in the control plane that is managed by cellular networks, and send/receive data into data planes based on decisions made by the control plane. Control plane includes two layers: social and decision layers. The first one includes a variety of social software. The latter makes different types of rules for data transfer, such as which type of communications can be used for data transfer(DSRC, Cellular or WIFI etc.). To be detailed in the social layer, first, each node based on the beaconing messages that received from base stations, connects to the nearest base station and request their offered services. If that service is available in that base station, it is going to be joined to that group to get its required services and if not, the base station creates a new shell that provides this service to the users. This paper mentions some challenges, such as how to manage the specific interest of each node and how to satisfy diverse and different requests that are coming from one node. The other challenge is about different carriers for supporting different nodes. How can we integrate them to have an integrated network to manage and control them properly in the control plane? Security and privacy is another challenging issue in this architecture as well. Figure \ref{picture16} shows a better view about control plane and its building blocks.

\begin{figure}[htbp]
\includegraphics[width=0.5\textwidth]{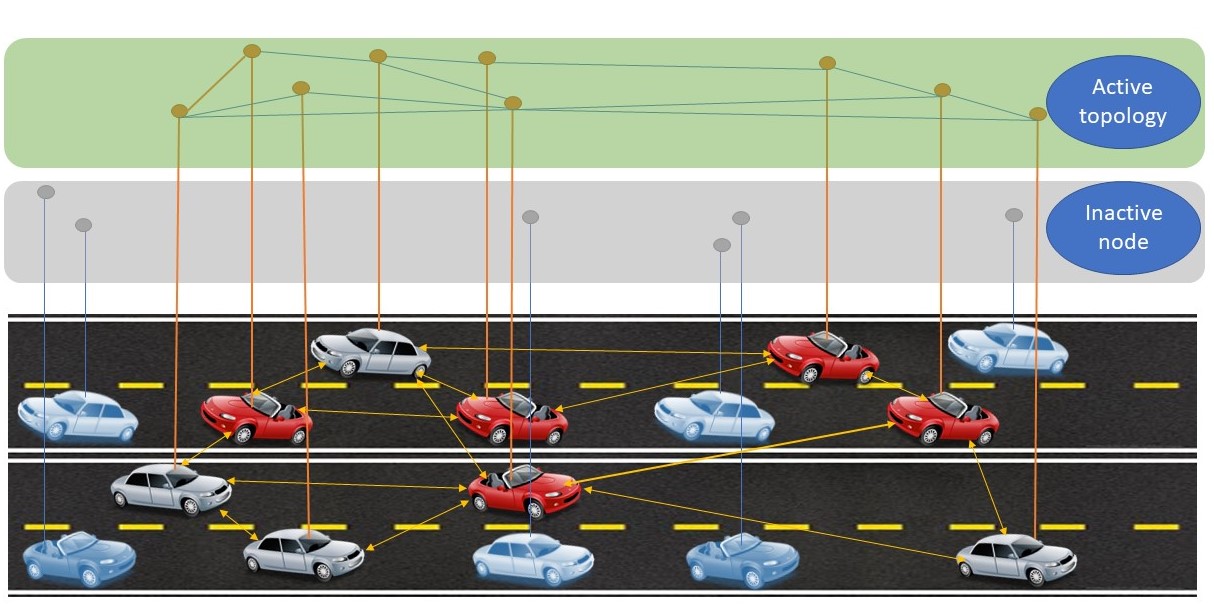}
\caption{Active topology extraction through having active and inactive users }
\label{activetop}
\end{figure}

Han et al.(2018) in \cite{han2018parallel} proposed an architecture based on two different systems: physical and virtual. Adding a new dimension to the vehicular network which is the cars social behaviour is useful but complex in most cases, as it's hard to predict human behaviour in emergency situations. Because of the unpredictability and uncertainty of both human and vehicular network behaviours, they proposed a new model that is highly correlated to inseparability and unpredictability features. In other words, all the systems that include human behaviour sometimes show unpredictable behaviours, and for modelling these complex systems, they used artificial intelligence to combine real objects and real society to use the full potential of AI from static to dynamic, offline to online and passive to proactive. There are a variety of factors that are given to an artificial intelligence system such as controllable factors and uncontrollable factors, user behaviour, and environmental parameters. This model has artificial societies, computational tests, and parallel implementation. 
\newline
Nitti in \cite{nitti2014adding} discussed a new term named SIoT \cite{atzori2011siot}\cite{atzori2011making}(short form of social internet of vehicles) which is a social connection between nodes and each node makes a new social relationship with respect to the established rules that are created by the owner. There are different types of social relationships between nodes, such as parental connection, which is defined by the manufacturer of the nodes that have been created at the same time by the same company and would not change over time. The co-location relationship is created between different nodes that exist in the same place and provide the same service. Some relationships between nodes can change over time. Ownership object connection is established between different nodes that have the same owner. Social object connection is created periodically or constantly for some reasons that are completely dependent on the owner's decisions. In SIoT architecture, there are three layers to provide services for the users. Nitti et al.\cite{nitti2014adding}, tried to create some novelties in the architecture of SIoT and named it SIoV. In SIoV, cars can create social connections and transfer data in order to improve the network efficiency in different metrics. In SIoV, each car can be considered as a source to give services to the users, such as traffic status and managing traffic. Also, based on the map that it has provided, it potentially knows when the user goes into a gas station or work etc. Finally, they provided an API to be added into SIoV to integrate cars into this model. One of its jobs is formatting the data in a predefined format, and then, send data to the servers. In their system, the management layer, security, and facility layer have been created to facilitate the connection and integrity of the vehicular network.
\newline
In \cite{arnaboldi2014cameo}, Arnoboldi et al. provides a middle-ware for social networks that have the mobility nature. They have two perspectives: social awareness and context awareness. They put together the user and her devices information , such as the name and application features on each device, her interest, her profile etc. With the information that is coming from other users, this model would be able to create new connections among users and new social cooperation between them. Data fusion of user and her device's information can be considered as a node. Each node is composed of three different components: local, external and social context. 
\newline
 \cite{iamnitchi2012social} is based on three different layers. The bottom layer includes current applications and potential future applications. The other layer is composed of different social sensors that are embedded in mobile phones and support different applications. Thus, this layer analyses a number of social signals and for the specific interaction, it transforms the domain. The other layer is responsible for fusing all the different information that is coming from social interactions between different nodes, to make social edges, and  data table about different social connections of the user with other nodes. Also, this layer is executed on the user device. Finally, when the aggregator fuses all the data which is personal, it sends the social relation graph or data table to another layer named SKS (short form of social knowledge service) for management and storing them. Each aggregator has different tasks as well. 
 \newline
 In \cite{rahim2019social}, Rahim et al. talked about different current challenges in the dynamic nature of the VANET and then proposed their multi-dimensions routing protocol solution. For data dissemination, there are two important factors: high mobility of the nodes and rapid topology movement that can be used for data routing between different nodes. Routing protocol based on location information might be useful in the highway, but challenging in urban scenarios. In this model, they consider the social features of each node such as how much it has a social connection with other nodes. Node centrality and activeness in routing messages are other features that are also taken into consideration. This architecture not only takes the current information of each metric, but also, they rely on historical data to make a better decision. The Store and carry approach is used by nodes to carry the messages between different or the same communities. Each community is established based on the same similar interests between nodes. When one node wants to send information to another node in the same community or different communities, they consider node centrality degree. Apart from node centrality, the degree level of acquaintance and activeness of nodes are important as well. Finally, they evaluated their model in terms of end-end delivery and packet delivery ratio in terms of speed and number of nodes, which showed better performance compared to two different protocols namely AODV \cite{chakeres2004aodv} and GPSR \cite{karp2000gpsr}.
 Paranjothi et al.(2016) in \cite{paranjothi2016mavanet} defined the connection between vehicles based on social networks, in which each car can connect to the other cars through social networks. Moreover, they can connect to other cars through using RSU and each RSU is directly connected to base stations to get the necessary information about each vehicle, in order to give them permission to be connected together. If a vehicle as a sender wants to connect to another vehicle as a receiver, it first evaluates the identity and current location of the receiver and then shapes a primary connection with the receiver. It's worth to mention that, messages in primary connections are encrypted between them in the shape of QR code and if authentication level between them was passed successfully, the encrypted message would be decrypted in the receiver side. Each node in a network, is differentiated from other nodes with email, username, id etc. QR or Quick Response Code is one of the encryption methods that save the data in the shape of dots, and it is fast in terms of encryption and decryption. The smallest version has 21*21 matrix size, while 177*177 is the largest version of QR code. They evaluated their approach in different metrics, such as the probability of dropping message, the delay time of the packets to be received in the receiver side, and the required time for both decryption and encryption in QR algorithm, which shows it will be affected drastically with the increasing number of users in all these metrics. Figure \ref{activetop}, demonstrates an overall view of active topology extraction from social networks.
\newline

Mahdi et al.(2017) in \cite{mahdi2017grouping} discussed different grouping such as classification and clustering which are out of the VANET networks. For example, the main technique in grouping nodes in the classification model is finding the nature of each group based on the similarity that they can have with each other. If the nature of each group would not be clear, instead of using a supervised technique, unsupervised techniques are going to be used for grouping them. We can have different techniques in using clustering techniques such as Euclidean or hamming distance measures \cite{sharma2014methods}. In clustering, as its name shows, each new member is going into a cluster in which their members have more similarities with that new member. These kinds of techniques are not operable in VANET, mainly because of having node mobility, as there is no defined stable topology. Furthermore, they tried to define two different types of grouping in VANET networks. The first one can be defined as a formal group, in which each member has a more stable connection with its group peers.  In the casual groups, cars can be grouped together based on the temporary types of behaviours they show. There are some factors that can be effective in casual grouping such as the density of the vehicles on the roads, as it's not equal in different parts of the city such as rural or urban areas. Finally, they used GrapStream for graph analysis and NetLogo \cite{tisue2004netlogo} for agent simulation.

In \cite{gainaru2009realistic}, Gainaru et al.(2009) presented a simulator named VNSim, which is a realistic simulator and has been designed in Java programming language. It includes two systems: it takes the behaviour of each car, and model a network based on macroscopic and microscopic movements of the cars. There are four types of modes for drivers: driving freely, braking, approaching, and following. The topology of the network based on cars can be gained by the drivers’ modes and personalities such as calm personality, normal or angry personality. This model can help the drivers to change their lane while driving, and have a traffic control. When it reads the points of entry and exit and then connects these two points together, it shapes a macroscopic view, which is helpful while the number of vehicles is small. However, this macroscopic view cannot be executable when the number of vehicles grows, which leads to the inability of this simulator for making the right topology between entry and exit points for each driver. For evaluating the behaviour of the cars in the context of sociability, they use two different creators for synthesizing social networks: Girvan-Newman and Caveman Model \cite{watts2004small}, which connects the cars which have the same final destination.
\newline
In \cite{liu2012exploring}, they collect data from moving cars in two different locations: shanghai and San Francisco cities to shape a vehicular network. They defined the social relationship between the vehicles, based on the number of encounters that they have. If the number of encounters is high, they have a high probability to be friends in the context of social networks. Every 10 seconds, they calculate whether they are still in a close connection with together or not. Then, they analyzed the social features of the network by testing some properties, such as clustering, small world phenomenon, and the distribution type of vertices’ degrees. The number of nodes decreased, when the level of degree between nodes is larger than 1002 which creates a straight line in the chart. Moreover, if the number of nodes is large in the network, there is more probability to have more encounters between the vehicles compared to the time that the number of nodes is not large enough. In their clustering coefficients, they showed that in both SUVnet-trace and Cabspotting \cite{eichler2007performance}, we have a higher degree compared to random graphs.

Huang et al.(2014) in \cite{huang2014social} proposed two different trust models in VANET namely: entity-oriented trust approach which mainly focuses on the level of trust between the cars and trust degree evaluation between them, and information-oriented trust approach which tells us about the trustworthiness of messages, which are transmitted in VANET. In other words, the first approach focuses on the cars to construct the trust, while the latter one addresses the trust level of messages which are transferred between the vehicles. One important challenge that exists in VANET is taking decisions based on received messages by one node, directly affects the decision that are made by other neighbouring nodes. For instance, if node 1 receives 5 different messages about car accidents in one specific location and 3 of these messages have been sent from malicious nodes, the decision that is being made by node 1 has an effect to its neighbouring nodes. So, for making the right decision based on the received messages by each node, if node 1 receives 2 right and 2 wrong messages about car accidents, it can evaluate the source of each message in terms of how much sender node is close to the event? Is it a direct observer node, or multi-hop observer node about the event?
\newline
Alam et al.(2014) in \cite{alam2014vedi} presented an architecture named VeDi, in which every driver is represented as a node and the connection between them is considered as a link. It includes five different sections: messages type based on DSRC, Onboard Unit (OBU), Roadside Unit (RSU), home-based unit(HBU), Cloud for VeDi and user interface for VeDi.  Each HBU has a data manager who collects data from OBU, and then, send them to the cloud. clearly, HBU is the intermediate nodes to connect OBUs to the cloud. User interface can be divided into different sections, such as social graphs, friends, different routes, and so on. 
In \cite{hu2012vssa}, Hu et al.(2012) propose a model to support different types of services for the users in VANET. It has two layers: the service and application layers. The application layer is accessible through end-user devices through a specific interface. With having service-oriented software models, we can run different applications for transportations scenarios.  In their service layer, they have three different parts namely: Aframe, social, and management services. Aframe is a framework that combines AmbientTalk \cite{dedecker2006ambient} and software agent model in a framework that has different layers and leads to the complexity reduction in the current communications among the nodes in the ad-hoc network. To test their model, they used REST \cite{richardson2008restful} services instead of SOAP \cite{box2000simple}(because of the high cost), and then used 8 people who have Android phones and tablets, which have been equipped with 802.11n wireless network. Each person sends a message every 50 seconds to different nodes when they are moving from one place to another at the speed of 20 km/h, and it showed that all the massages were received to the destination successfully.
\newline
In \cite{raza2018social}, Raza et al.(2018) proposed a communication model for ITS in 5G. The main goal of this model is to enable cars to send vital messages about different subjects, such as road status, congestion, and accident alerts to everything(V2X) which is run in the MEC server. The network component consists of different eNodeBs, different modules for network core, and SDN switches. There are some fog servers that are implemented near the edge devices, which offer different services for the end-users. The other module is V2X AS, which has the ability to run different applications, such as the applications in ITS field, social applications, and eMBMS applications. Also, Software-Defined Network (SDN) is responsible for data managing and controlling in two phases: data plane and control plane. Another server in this model is Social IoT. It's responsible to make the integration of IOT devices easier, and activates different actions, based on application rules. Finally, there is a cloud that maintains all the information about the network, and has an abstract view of different parts of the network. It stores all the data about the Social Internet of Vehicles (SIoV). Cars can create a social relation with their neighbours and share the data that is useful, such as road status. One of the useful things in this architecture is, having awareness cooperation between different entities listed from road users to roadside infrastructures, MEC servers etc. For traffic management, this model uses data that is coming from Google, which is originated from end-user devices and different applications that are installed in them. For their performance evaluation, they used VSimRTI (short form of Simulation Runtime Infrastructure) \cite{schunemann2011v2x} which has a number of powerful simulators in itself.
\newline
In \cite{ning2017cooperative}, they discuss a cooperative and quality-aware model for SIoV (Social Internet of Vehicles). User interest can be seen as a connection map between devices and social connections in the real world. Only similar interests between the nodes can form a group. Each user in the network can show two different types of behaviours, which is normal (providing the correct information about one event) and abnormal (providing false information and evaluation about one event). Social relationships can be defined in 4 different types: parental relationship, social relationship, co-location and co-work relationships. 
\newline
In \cite{zia2019towards}, they consider both the user and her car as one entity. The main goal of this model is evaluating different Social internet of Vehicles in offering services to other nodes, especially when there is no choice by themselves. Each car completely knows about different plans that is made by its owner and different good or bad experiences about different resources. We can have different assumptions based on the user activity. For example, a specific user can request the same services at different times from the servers, in which this service can be high or low in terms of QoS in different servers. They considered different hypotheses to evaluate their performance through the network. The main goal of this model is to help the entities to choose the best resources for getting the services based on the good and bad experiences that the users have in choosing the resources. They designed an overall system flow, in which they have a matrix plan with having three activities, which are chosen randomly. Each car has a certain level of expectation from the resources, and if the QoS of visited resources do not meet its requirements (or based on the historical data, it concludes that its expectation cannot be satisfied from visited resources), they completely excluded from the offered resources, and then, the recommendation system starts working in the network. They defined different strategies for plan execution. In their simulation results, they showed strong social ties with timing effects.

Maglaras et al.(2016) in \cite{maglaras2016social} grouped vehicles into different clusters, and all of these clusters can build the main structure of the networks. For building a cluster based on the similar interests, each cluster has a fixed member that includes cluster head and mobile members. Each fixed member has all the information about the other member. Each node can make a P2P connection based on the degree of similar interest between peers. After joining into a group, it submits its information to the cluster head. Through using this structure, each node can get its required files from cluster heads or other mobile peers. For getting the file from neighbouring peers, it takes the amount of connection between peer nodes to evaluate path lifetime\cite{wang2014real}\cite{marquez2014breaking}. If one node has a long connection time with other nodes, it acts as a forwarder to send the request to other neighbouring nodes, otherwise, it just checks that whether it has the requested file or not. Compared to the other methods such as Gnutella \cite{cheng2011infotainment} and an Interest-based P2P scheme, this method works much better, as these two methods pick other nodes for forwarding the messages randomly.
\subsection{Recent advancements in VANET using 5G, SDN and Fog}
\begin{figure}[htbp]
\includegraphics[width=0.5\textwidth]{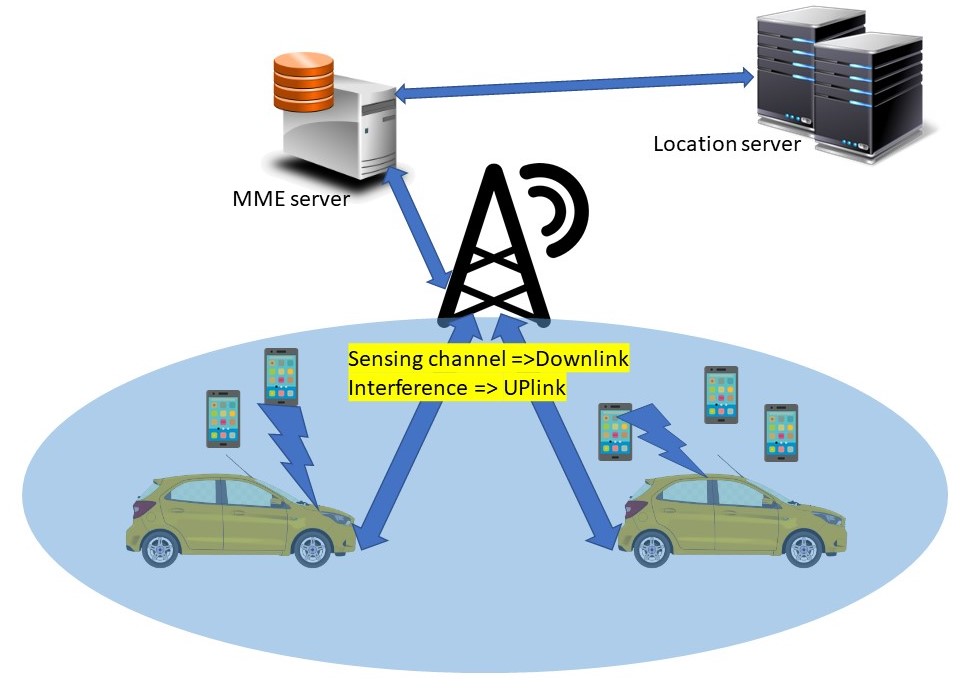}
\caption{Network Scenario }
\label{picture18}
\end{figure}

In \cite{mumtaz2015cognitive}, Mumtaz et al.(2015), first talked about two different modes that LTE-A works on that, FDD (short form of frequency division multiplex) and TDD (short form of time-division multiplex). TDD is helpful for the systems that are asymmetric. TDD uses uplink and downlink (UL and DL) though giving  time slots to senders in a current bandwidth. LTE-A-A FDD uses both downlink and uplink spectrums, because LTE-A FDD just used downlink band and the uplink band is not being used by it. This paper \cite{mumtaz2015cognitive}, uses LTE-AA UL. Suppose that each node has been equipped with LTE-A technology and there is one node LTE-A eNB, which is not near V2V radio communication type. For avoiding interference in the channel, on the receiver side, there is a threshold for tolerating temperature due to having a number of interferences. If the sender meets this requirement and the amount of interference level is below the threshold, they can send their data. V2V network has the capability to get the path loss among its location and LTE-A-A eNB. The information that is achieved by V2V for sensing path loss is important for LTE-A eNB in order to control transmission power in V2V for dealing with interference.
\newline
 For having cooperation between V2V and LTE-A-A, having a rule for avoiding interference is really important. All the resources in the network are being managed by eNB. Also, LTE-A eNB can receive interference, if V2V sends data in uplink in the same band. In their research, they proposed a simple approach for managing interference. Also, they supposed that there is one eNB that is located in the cell centre, and there are two different types of users: cellular and V2V users. The first one is named primary and the latter one is named secondary users. LTE-A eNB calculates the maximum amount of transmission power that each V2V user can use in order to not interfere with eNB transmission power. For getting the maximum amount of transmission power for V2V communications, first, there is a node to explore the path loss among V2V nodes location and eNB. As there is a strong relation between UL and DL, they first sense DL signals, for finding the good opportunities to exploit UL band in LTE-A. Moreover, the antenna of eNB is located in a high position which leads to finding DL signal much easier than UL signals which are sent through different users from different locations. Another benefit of sensing DL link is having a synchronized manner which makes the detection process easier through using the different features of cyclostationary in the LTE-A signal. Figure \ref{picture18} illustrates their scenario.

\begin{figure}[htbp]
\includegraphics[width=0.5\textwidth]{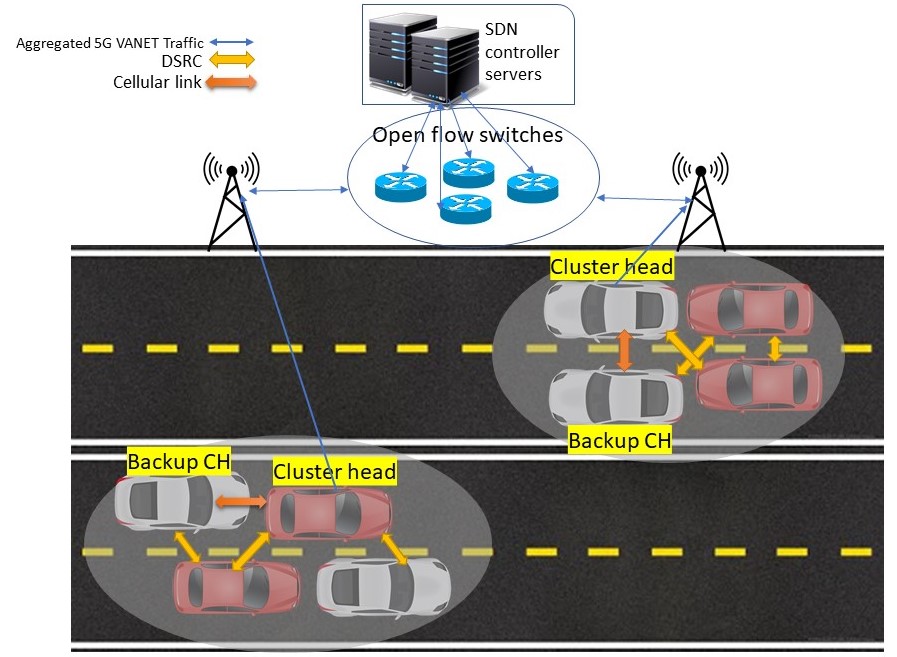}
\caption{Having Clusterhead and Backup Cluster head }
\label{picture19}
\end{figure}

In \cite{cheng2018big},  Chenge et al.(2018) proposed a new approach for detecting NLoS (Non-Line of Sight) through V2V. It helps the system to allocate resources and find relay nodes at the right time to the nodes that are under the NLoS condition. They divided their model into two parts: Firstly, they collect data set about V2V connection type measurement through using two cars that have been equipped with OBU that have DSRC module. Moreover, they employed two cameras in each receiver and sender in front and back rears. These camera controls and records the process in order to have offline analysis. They collect data in three different scenarios: urban, highway, and suburban. The second part of their model is using machine learning algorithms for detecting NLoS parts in the city. They used two well-known algorithms namely NV (Naïve Base) and SVM (Support Vector Machine). 
\newline
In \cite{cheng20175g},   Chenge et al. used SDN, cloud, and fog nodes in their architecture. Based on the different requirements that are coming from different vehicles, this framework can decide about which framework layer is responsible to analyze or store received data, which gives us flexibility in making decisions in different scenarios. In addition for self-driving, cars can have access to the data to learn how to imitate driving safely. The other key factor is, having a heterogeneous architecture and all different entities are cooperating with each other. Having an open standard to help them for coexistence seems necessary. Also, through having an open standard, this framework can better support the network in terms of keeping an intelligent connection between  vehicles. For example, macrocell BS keeps cars control messages to assure seamless connectivity in a highly mobile network. On the other hand, huge data exchange is done in microcells and femtocells through D2D technology. In this regard, we can guarantee high reliability and lower latency. 
\newline
In \cite{duan2016sdn}, Duan et al. talked about SDN-5G VANET. In this architecture, they used SDN for coordination among heterogeneous nodes. In their architecture, each car can get its required services from eNB through its cluster head. IEEE 802.11p is the communication type between group members in each cluster, that can save cellular communication resources. There is a backup cluster head that saves signalling messages from the current cluster head and prepares itself to take the responsibility of the current cluster head in emergency situations. SDN control plane is analyzed and evaluated to explore the network performance in terms of doing handover without too much delay. Also, the controller has a global view of the network, and clustering is done whenever it's needed.
\newline
In \cite{qi2018sdn}, Qi et al.(2018) showed another clustering technique in the 5G-VANET system by using SDN. SDN has different parts: application, control, and data planes.  In the control plane, they can monitor and control the global map of the network and provide a seamless management through the whole network in Southbound Interface (SBI). Based on the historical data, they can predict the location of the vehicles. Also, the rules are sent to the cars by cellular communication links, and to the Base station by fiber links after doing clustering.Also, Data plane includes a base station and vehicles that are equipped with OBU.  After the cars send their related data, such as their IDs and their path to BS, They use a semi Markov model for location prediction of the cars and considered historical data and temporary time probability as inputs, to predict the social pattern of the cars.  Furthermore, they compared their algorithm with two other well-known algorithms  in terms of cluster lifetime vs vehicle velocity, (LID \cite{gerla1995multicluster} and MPBC \cite{ni2011mpbc}), which showed better results. In figure \ref{picture19}, 5G VANET integrated architecture has been depicted.

In \cite{liu2017high}, Liu et al.(2017) talked about having 4 layers in their model. The first layer is a sensing layer that senses the environment through OBUs and RSUs. Sensors that are implemented in the cars are responsible to sense the speed, car direction, related information about its surrounding environment, temperature, and even humidity. They are considered active data. RSU is responsible for knowing the number of cars in its surrounding environment and their plate numbers that called passive data. In the communication layer, they used MIMO (multiple input multiple output) and cognitive radio technology for getting 1.2 Gb/s. OpenFlow protocol is used by SDN. Moreover, for better control in traffic management, they have 3 different components such as vehicle localization, which are helpful when we do not have access to GPS, traffic data processing (pre-fetching, synchronizing, and storing data), and traffic service prediction (using deep learning). In their performance evaluation, they took different accidents in different locations and compare their model in the presence of police and absence of the police in rush hours and non-rush hours which showed the shorter ambulance arrival time compared to current models.
Storck et al.(2019) in \cite{storck20195g}, discuss a framework named 5G V2X ecosystem that uses 5G and SDN. In their architecture, they have a Baseband Unit that has a connection with the SDN controller. The Baseband unit connects to the 5G Core Network by using a backhaul link. In this architecture, they have network slicing that separates the control plane and user plane. There are four communication types: V2I, V2N, V2V, and V2P. In the core network, there are different entities that are shared between different slices. Through slicing different services, we can conclude that, each car can have different slices based on their needs. For example, one car can have two slices. One of them is an autonomous driving slice and the other one is for entertainment.

In \cite{din20195g},  Din et al.(2019) used 5G architecture that has different hierarchical layers. The first layer is the sensing layer, which is responsible for sensing the environment by RSU and cars. The other layer is the relay layer, which has been established to help a long communication time between devices. Connected RSUs can create a relay layer. The relay layer is responsible for sending the received data to its upper layer. As we have mobility in our architecture, cars send related information about their speed, source, destination, type of message, and so on frequently. The third layer is the convergence layer and massive MIMO supports this layer. MIMO has a high antenna that sends the packets to the BS.  Their proposed SDN based architecture is performed in 5 levels. Also, the data is collected from different entities and then, are being sent to the processing level. They process real-time data through different methods such as Spark \cite{zaharia2010spark}, GrapheX, and Hadoop \cite{kaushik2010greenhdfs} ecosystem.Also, they divide the tasks into sub-tasks. The last layer is the application layer that can be divided into two important layers: NDN (short form of Named Data Network), and event controlling management.

In \cite{aadil2018clustering}, each node has been equipped with a 5G interface. They used the 5G spectrum for backup infrastructure. Clustering in a heterogeneous network helps to have robust communications. There are some clustering techniques, such as a nature-inspired swarm intelligence algorithm \cite{abraham2008swarm}, which is energy consuming and is helpful when the energy-consuming at a high level is not a big deal. Locations of the vehicles are usually gained by GPS. But, in 5G we have an entity named mobility management that includes the location of the users, and it's connected to eNB. They minimize the number of clusters in order to have stability in the network. For overhead minimization, each member in the cluster connects to the Cluster Head (CH) and only CH sends the information to CMs(Cluster Members). They used swarming techniques of dragonflies for clustering. In this technique, each node is considered as one solution and a swarm is a group of nodes or solutions. Although this technique is energy consuming and has high complexity, it's one of the best algorithms for clustering in IoV. 
\newline
In \cite{duan2017sdn}, Duan et al. proposed a cluster-based architecture in the heterogeneous network, it has different layers namely small cells, macrocells, BS, and AP (Access Points). Instead of having a controller in different entities, such as BS or AP, there is a centralized controller that is equipped with OpenFlow protocol. As a result, all the clustering, authentication, traffic rules, and different new functions are executed in the SDN controller. The information about all cells in the SDN controller can be updated proactively (periodically) or reactively (based on request). When the vehicle clustering is formed, beamforming transmissions in the link between each cluster head and the base station will be created. Beamforming techniques are run in the area with a high load of the cars. They showed two different ways of beamforming. The first one just covers Cluster Head to avoid intra-beam interference and in the latter one, beamforming covers all the clusters, not just Cluster Head. The second one is preferable in the environment with high mobility, as we have a wider beam.
\newline
In \cite{vladyko2019distributed}, Vladyko et al.(2019) proposed a system that has two main components, SDN and MEC. It's important to note that, both physical and MAC layers of IEEE 802.11p are derived from another standard named IEEE 802.11a. IEEE 802.11p has some benefits and some challenges for safety and non-safety applications. For dealing with these challenges, another standard has been developed named IEEE 802.11bd which can offer higher throughput (double than IEEE 802.11p). The supported speed is about 500km/h, and it supports more coverage area than 1km. However, it has compatibility problems with current systems. In this architecture, there are three different communication types: V2V, V2X, V2I. there are 4 different layers in this system namely: the physical layer that OBU and RSU are placed in this layer. The second layer is our MEC devices that they are heterogeneous cloud servers. They can be divided into two servers: Micro cloud and mini cloud servers. Micro cloud has limited capabilities in terms of computation and storage. A group of micro clouds are directly connected to a mini cloud server that has more computation and storage capabilities. The third layer has two different planes: the data plane and the control plane in which the control plane is responsible for managing the network in a distributed manner. The fourth layer shows the application servers and the core network gives the interface to connect the third layer to the fourth layer. 

In \cite{ge2016vehicular}, Ge et al.(2016) proposed an architecture for VANET that has 2 tier cellular networks. The first one is macrocell BS, that is responsible for managing and controlling the data belongs to cars and small cell BS. The latter one named small cell BS that sends the packets to the cars. Control plane can be formed by macrocells, and small cells can shape user plane. The signals that are carried by a physical downlink, belong to the control plane and the data belongs to users, are carried by the User plane. They showed the cooperation between small cell Base stations that can send the data to the related vehicles. In their performance evaluation, they showed that cooperation between small cells can improve the VANET communication capacity and handoff rates.

In \cite{alghamdi2020novel},  they clustered all the cars with the same path or similar path to have a dynamic network. Optimal forwarders support V2V and optimal devices forward D2D communications. People as pedestrians can act as a gateway for V2I. In order to satisfy the transmission of safety messages, the best transmitter is chosen to make the dissemination performance better. For safety message dissemination, a source that might be an accidental vehicle, sends their related data to the Cluster Head and CH chooses the best disseminator through different factors. This method can increase QoS in the 5G VANET environment.
\newline
In \cite{khan2019two}, Khan et al.(2019) showed that for direct V2V communications, each car supports 2 different modes. The difference between them is about allocating resources. In the first mode, resources are being allocated by cellular networks. On the other hand, in another mode, the vehicles pick the resources without getting help from cellular networks through distributed scheduling schemes. In their proposed scheme, they have two-level clustering approaches, where cars are being grouped into two different layers. There are some factors such as link reliability, the cars velocities, and k connectivity of the cars that can form the first layer through using the fuzzy logic method. The second cluster is obtained from the first cluster formation. This can be an optimal gateway for data distribution. They used \cite{konar2013deterministic} which is an improved Q-learning (IQL) in which, only the best actions are put in the table. They showed that this method is better than classical Q learning (CQL) \cite{wu2018cluster} method in terms of low computational costs and the number of tuned gateways in LTE Base Stations.

In \cite{song2014handover}, Song et al.(2014) proposed a two-level clustering model for 5g VANET communications. First, in order to have less interference and less congested channel, they split the network into  control plane and date plane which is supported by macro eNB for having a reliable channel for controlling information. On the other hand, phantom eNB is responsible for managing and controlling user traffic data with a higher spectrum inside the small cells. Both control and user plane have the same LTE frequency domain. But, this frequency domain is divided between control and user planes, and only data generated by the users are carried by the user plane. Handover is challenging when, we have inter macro cell handovers compared to normal LTE networks. They redesigned the activation of handover with the GM model (one of the factors that are important for making handover decisions is about the quality of received signal) and handover signalling flow. They showed that not only they can predict when to trigger handover, but they have an improvement in inter macro eNB handover mechanism.
\newline
In \cite{soua2018multi}, Soua et al.(2018) showed a multi-level SDN with fog computing in 5g VANET networks. SDN architecture in this architecture has a hybrid structure in which, a car can be a local controller inside the SDN structure. There are backup SDN agents as well, to help the primary SDN in emergency cases. Through having fog and SDN in this system, they have a distributed controller. They tried to improve the idea of vehicles as infrastructures in \cite{hou2016vehicular}. This system consists of 2 sub-models: the first one has resided in the central SDN controller named permanent cloud, and the second one is temporary fogs that make a local decision by local SDN controller. It has some benefits, such as forming a fog by cars, and each fog can act as a sub SDN, and decreases the traffic that goes to the cloud. Moreover, each sub SDN can support real-time applications and analyses received data in real-time. 
\newline
In \cite{ning2019vehicular}, they use the benefits of fogs and vehicular communications. It has three layers: cloud, cloudlet, and fog. The Cloud layer consists of a traffic management controller and a trusted third authority that controls and manages the traffic of the whole city. For avoiding the data overload in the traffic management controller that sent by vehicles, this system has individual rewards and having fairness in the network that is controlled by the trusted third authority. They divided the city into different regions, and each cloudlet is responsible for one region. The third layer is the fog layer that has vehicles in the coverage area of each RSU. Fog nodes can be formed by moving or parked cars and the sensed information can be sent to the RSU for further processing. 

In \cite{khattak2019integrating}, they used both cloud and fog concepts for time-sensitive applications in the VANET. A vehicular fog network has three different layers: vehicular network, vehicular fog, and service provisioning layers. Vehicular network layer is consisting of vehicles that are equipped with OBU. These cars as the author said, have some basic capabilities such as storage and processing and can connect to the edge devices for further data processing. Vehicular fog layer includes RSUs that are installed in important spots such as junctions or parking spots and these RSUs have consistent links with cloud and connection with the cars. The service provisioning layer offers huge capabilities and can communicate with fog nodes and orchestrates the requests that are coming from consumers to the fog nodes.

In \cite{bitam2015vanet}, they proposed a system that is composed of two sub-layers: permanent and temporary clouds. Permanent cloud has all the features of traditional clouds, such as high capabilities, virtual machines, and vehicles, RSU, and other entities  can exploit them. The temporary cloud has all the available resources ranging from stationary to mobile entities. So, all of these resources are accessible from end-users. We have three different main layers in this architecture: Client layer which includes end devices such as smartphones, laptops and cars. Another layer is the communication layer, which aims to ensure the reliable communication between clients in the lower layer and the VANET cloud in the upper layer.  The third layer is a cloud layer that has the same functionality as the traditional stationary cloud layer, and all the temporary resources in temporary Cloud belong to the cars and users in the cars.

\begin{table*}[hbt!]
\caption{Comparison between papers in 4 different metrics}
\begin{tabular}{|c|c|c|c|c|}

\hline
\textbf{Paper} &
  \textbf{\begin{tabular}[c]{@{}c@{}}Incentive\\ techniques\end{tabular}} &
  \textbf{\begin{tabular}[c]{@{}c@{}}Handover\\ Management\end{tabular}} &
  \textbf{\begin{tabular}[c]{@{}c@{}}Data \\ Offloading\end{tabular}} &
  \textbf{\begin{tabular}[c]{@{}c@{}}Resource\\ Allocation\end{tabular}} \\ \hline
\textbf{\cite{nobre2019vehicular}}  & \textbf{}  & \textbf{}  & \textbf{*} & \textbf{}  \\ \hline
\textbf{\cite{luo2011geoquorum}}  & \textbf{}  & \textbf{*} & \textbf{*} & \textbf{}  \\ \hline
\textbf{\cite{khattak2019toward}}  & \textbf{}  & \textbf{*} & \textbf{*} & \textbf{}  \\ \hline
\textbf{\cite{sun2016edgeiot}}  & \textbf{}  & \textbf{}  & \textbf{*} & \textbf{}  \\ \hline
\textbf{\cite{luo2018cooperative}}  & \textbf{}  & \textbf{*} & \textbf{*} & \textbf{}  \\ \hline
\textbf{\cite{khan20185g}}  & \textbf{}  & \textbf{}  & \textbf{}  & \textbf{*} \\ \hline
\textbf{\cite{tomovic2017software}}  & \textbf{}  & \textbf{}  & \textbf{}  & \textbf{*} \\ \hline
\textbf{\cite{cruz2018sensingbus}}  & \textbf{}  & \textbf{}  & \textbf{*} & \textbf{}  \\ \hline
\textbf{\cite{ning2018mobile}}  & \textbf{}  & \textbf{}  & \textbf{*} & \textbf{*} \\ \hline
\textbf{\cite{gao2017fog}} & \textbf{}  & \textbf{}  & \textbf{*} & \textbf{}  \\ \hline
\textbf{\cite{shah2019vfog}} & \textbf{}  & \textbf{}  & \textbf{}  & \textbf{}  \\ \hline
\textbf{\cite{lai2018fog}} & \textbf{}  & \textbf{}  & \textbf{}  & \textbf{*} \\ \hline
\textbf{\cite{ansari2018cloud}} & \textbf{}  & \textbf{}  & \textbf{*} & \textbf{}  \\ \hline
\textbf{\cite{xiao2019fog}} & \textbf{}  & \textbf{*} & \textbf{*} & \textbf{}  \\ \hline
\textbf{\cite{paranjothi2018dfcv}} & \textbf{}  & \textbf{}  & \textbf{}  & \textbf{*} \\ \hline
\textbf{\cite{soua2018multi}} & \textbf{}  & \textbf{}  & \textbf{}  & \textbf{}  \\ \hline
\textbf{\cite{khattak2019integrating}} & \textbf{}  & \textbf{}  & \textbf{}  & \textbf{}  \\ \hline
\textbf{\cite{lobo2017solve}} & \textbf{}  & \textbf{}  & \textbf{}  & \textbf{}  \\ \hline
\textbf{\cite{brennand2017novel}} & \textbf{}  & \textbf{}  & \textbf{}  & \textbf{}  \\ \hline
\textbf{\cite{liang2019extremely}} & \textbf{}  & \textbf{}  & \textbf{}  & \textbf{}  \\ \hline
\textbf{\cite{li2016veshare}} & \textbf{}  & \textbf{*} & \textbf{}  & \textbf{*} \\ \hline
\textbf{\cite{han2018parallel}} & \textbf{}  & \textbf{}  & \textbf{}  & \textbf{}  \\ \hline
\textbf{\cite{nitti2014adding}} & \textbf{}  & \textbf{}  & \textbf{}  & \textbf{}  \\ \hline
\textbf{\cite{arnaboldi2014cameo}} & \textbf{}  & \textbf{}  & \textbf{}  & \textbf{}  \\ \hline
\textbf{\cite{iamnitchi2012social}} & \textbf{}  & \textbf{}  & \textbf{}  & \textbf{}  \\ \hline
\textbf{\cite{rahim2019social}} & \textbf{}  & \textbf{}  & \textbf{}  & \textbf{}  \\ \hline
\textbf{\cite{paranjothi2016mavanet}} & \textbf{}  & \textbf{}  & \textbf{}  & \textbf{}  \\ \hline
\textbf{\cite{mahdi2017grouping}} & \textbf{}  & \textbf{}  & \textbf{}  & \textbf{}  \\ \hline
\textbf{\cite{gainaru2009realistic}} & \textbf{}  & \textbf{}  & \textbf{}  & \textbf{}  \\ \hline
\textbf{\cite{liu2012exploring}} & \textbf{}  & \textbf{}  & \textbf{}  & \textbf{}  \\ \hline
\textbf{\cite{huang2014social}} & \textbf{*} & \textbf{}  & \textbf{}  & \textbf{}  \\ \hline
\textbf{\cite{alam2014vedi}} & \textbf{}  & \textbf{*} & \textbf{}  & \textbf{}  \\ \hline
\textbf{\cite{hu2012vssa}} & \textbf{}  & \textbf{}  & \textbf{}  & \textbf{}  \\ \hline
\textbf{\cite{raza2018social}} & \textbf{}  & \textbf{}  & \textbf{}  & \textbf{}  \\ \hline
\textbf{\cite{ning2017cooperative}} & \textbf{}  & \textbf{}  & \textbf{}  & \textbf{}  \\ \hline
\textbf{\cite{zia2019towards}} & \textbf{}  & \textbf{}  & \textbf{}  & \textbf{}  \\ \hline
\textbf{\cite{maglaras2016social}} & \textbf{}  & \textbf{}  & \textbf{}  & \textbf{}  \\ \hline
\textbf{\cite{luo2019software}} & \textbf{}  & \textbf{}  & \textbf{}  & \textbf{}  \\ \hline
\textbf{\cite{cheng2018big}} & \textbf{}  & \textbf{}  & \textbf{}  & \textbf{*} \\ \hline
\textbf{\cite{cheng20175g}} & \textbf{}  & \textbf{}  & \textbf{}  & \textbf{*} \\ \hline
\textbf{\cite{duan2016sdn}} & \textbf{}  & \textbf{}  & \textbf{}  & \textbf{}  \\ \hline
\textbf{\cite{qi2018sdn}} & \textbf{}  & \textbf{}  & \textbf{}  & \textbf{}  \\ \hline
\textbf{\cite{liu2017high}} & \textbf{}  & \textbf{}  & \textbf{}  & \textbf{}  \\ \hline
\textbf{\cite{storck20195g}} & \textbf{}  & \textbf{}  & \textbf{}  & \textbf{}  \\ \hline
\textbf{\cite{din20195g}} & \textbf{}  & \textbf{}  & \textbf{}  & \textbf{}  \\ \hline
\textbf{\cite{aadil2018clustering}} & \textbf{}  & \textbf{}  & \textbf{}  & \textbf{}  \\ \hline
\textbf{\cite{duan2017sdn}} & \textbf{}  & \textbf{}  & \textbf{}  & \textbf{}  \\ \hline
\textbf{\cite{vladyko2019distributed}} & \textbf{}  & \textbf{}  & \textbf{*} & \textbf{*} \\ \hline
\textbf{\cite{ge2016vehicular}} & \textbf{}  & \textbf{}  & \textbf{*} & \textbf{}  \\ \hline
\textbf{\cite{alghamdi2020novel}} & \textbf{}  & \textbf{}  & \textbf{}  & \textbf{}  \\ \hline
\textbf{\cite{khan2019two}} & \textbf{}  & \textbf{}  & \textbf{}  & \textbf{*} \\ \hline
\textbf{\cite{song2014handover}} & \textbf{}  & \textbf{}  & \textbf{}  & \textbf{}  \\ \hline
\textbf{\cite{soua2018multi}} & \textbf{}  & \textbf{}  & \textbf{}  & \textbf{}  \\ \hline
\textbf{\cite{ning2019vehicular}} & \textbf{*} & \textbf{}  & \textbf{}  & \textbf{}  \\ \hline
\textbf{\cite{khattak2019integrating}} & \textbf{}  & \textbf{}  & \textbf{}  & \textbf{}  \\ \hline
\textbf{\cite{bitam2015vanet}} & \textbf{}  & \textbf{}  & \textbf{*} & \textbf{}  \\ \hline
\end{tabular}
\end{table*}

\begin{table*}[]
\caption{Comparison between papers in 15 different metrics.
U=Urban/H=Highway/R=Rural}
\centering
\resizebox{\textwidth}{!}{%
\begin{tabular}{|c|c|c|c|c|c|c|c|c|c|c|c|c|c|c|c|}
\hline
Paper &
  \begin{tabular}[c]{@{}c@{}}Heter.\\ support\end{tabular} &
  \begin{tabular}[c]{@{}c@{}}Infras\\ Need\end{tabular} &
  \begin{tabular}[c]{@{}c@{}}Geo.\\ Dist\end{tabular} &
  \begin{tabular}[c]{@{}c@{}}Loc\\ Awarness\end{tabular} &
  \begin{tabular}[c]{@{}c@{}}Ulrta\\ low\\ latency\end{tabular} &
  \begin{tabular}[c]{@{}c@{}}Mobility\\  support\end{tabular} &
  \begin{tabular}[c]{@{}c@{}}Real\\ Time\\ App\\ Support\end{tabular} &
  \begin{tabular}[c]{@{}c@{}}Large\\ scale\\ App\\ Support\end{tabular} &
  \begin{tabular}[c]{@{}c@{}}Multiple\\ IoT\\ App\\ Support\end{tabular} &
  \begin{tabular}[c]{@{}c@{}}Virtualization\\ Support\end{tabular} &
  \begin{tabular}[c]{@{}c@{}}Region\\ SPT\end{tabular} &
  \begin{tabular}[c]{@{}c@{}}Scalability\\ Support\end{tabular} &
  SDN &
  VSN &
  Cloud \\ \hline
\cite{nobre2019vehicular}  &  * & * & * &   & * & * & * &   &   &   & U   &   & * &   & * \\ \hline
\cite{luo2011geoquorum}  &                                                  & * & * &   &   & * &   &   &   &   & UH  & * &   &   & * \\ \hline
\cite{khattak2019toward}  &                                                  & * &   & * &   & * &   &   & * &   & U   &   &   &   &   \\ \hline
\cite{sun2016edgeiot}  & *                                                & * & * & * &   & * &   &   & * & * & UH  & * &   &   & * \\ \hline
\cite{luo2018cooperative}  & *                                                & * & * & * &   & * &   & * &   &   & UH  & * &   &   & * \\ \hline
\cite{khan20185g}  &                                                  & * & * & * &   & * &   &   &   &   & UH  & * & * &   & * \\ \hline
\cite{tomovic2017software}  & *                                                & * & * & * &   & * &   &   &   & * & UH  & * & * &   & * \\ \hline
\cite{cruz2018sensingbus}  &                                                  & * & * & * &   & * &   &   &   &   & U   & * &   &   & * \\ \hline
\cite{ning2018mobile}  & *                                                & * & * & * &   & * &   &   & * &   & UHR & * &   &   &   \\ \hline
\cite{gao2017fog} & *                                                & * & * & * & * & * & * &   & * &   & UH  & * &   &   & * \\ \hline
\cite{shah2019vfog} &                                                  &   & * & * &   & * &   &   &   &   & UH  & * &   &   & * \\ \hline
\cite{lai2018fog} & *                                                & * & * & * &   & * & * &   &   &   & UH  & * &   &   & * \\ \hline
\cite{ansari2018cloud} & *                                                & * & * & * &   & * & * & * &   &   & UH  & * &   &   & * \\ \hline
\cite{xiao2019fog} &                                                  & * & * & * &   & * & * &   & * &   & UHR & * & * &   & * \\ \hline
\cite{paranjothi2016mavanet} & *                                                & * & * & * &   & * & * &   &   &   & UH  & * & * &   & * \\ \hline
\cite{soua2018multi} & *                                                & * & * & * & * & * & * &   &   &   & UH  & * &   &   & * \\ \hline
\cite{khattak2019integrating} & *                                                & * & * & * &   & * & * & * & * &   & UH  & * &   &   & * \\ \hline
\cite{lobo2017solve} & *                                                & * & * & * &   & * & * &   &   &   & UH  & * &   &   & * \\ \hline
\cite{brennand2017novel} & *                                                & * & * & * &   & * &   &   &   & * & UH  & * &   &   &   \\ \hline
\cite{liang2019extremely} & *                                                & * & * & * &   & * &   &   &   &   & UHR & * &   &   &   \\ \hline
\cite{li2016veshare} & *                                                & * & * & * &   & * &   & * & * &   & UH  & * & * & * &   \\ \hline
\cite{han2018parallel} & *                                                &   & * & * &   & * & * &   & * & * & UHR & * &   & * &   \\ \hline
\cite{nitti2014adding} &                                                  & * & * & * & * & * & * &   &   &   & UHR & * &   & * &   \\ \hline
\cite{arnaboldi2014cameo} &                                                  & * & * & * &   & * &   &   &   & * & UHR & * &   & * &   \\ \hline
\cite{iamnitchi2012social} &                                                  & * & * & * &   & * & * & * & * &   & UH  & * &   & * &   \\ \hline
\cite{rahim2019social} &                                                  &   & * & * &   & * & * &   &   & * & H   &   &   & * &   \\ \hline
\cite{paranjothi2016mavanet} & *                                                & * & * & * &   & * & * &   & * &   & UR  & * &   & * &   \\ \hline
\cite{mahdi2017grouping} &                                                  &   & * & * &   & * & * &   &   &   & U   & * &   & * &   \\ \hline
\cite{gainaru2009realistic} &                                                  & * &   & * &   & * &   & * & * &   & UHR & * &   & * &   \\ \hline
\cite{liu2012exploring} &                                                  &   & * & * &   & * &   &   &   &   & U   &   &   & * &   \\ \hline
\cite{huang2014social} &                                                  &   & * & * &   & * & * &   & * &   & U   & * &   & * &   \\ \hline
\cite{alam2014vedi} & *                                                & * & * & * &   & * & * & * & * &   & UH  & * &   & * & * \\ \hline
\cite{hu2012vssa} & *                                                & * & * & * &   & * & * & * &   &   & U   &   &   & * &   \\ \hline
\cite{raza2018social} & *                                                & * & * & * & * & * & * &   &   &   & UH  & * & * & * & * \\ \hline
\cite{ning2017cooperative} &                                                  &   & * & * &   & * & * &   &   &   & UH  & * &   & * &   \\ \hline
\cite{zia2019towards} &                                                  &   & * & * &   & * &   &   &   &   & UH  & * &   & * &   \\ \hline
\cite{maglaras2016social} &                                                  & * & * & * &   & * & * &   &   &   & UH  & * &   & * &   \\ \hline
\cite{luo2019software} &                                                  & * & * & * & * & * & * & * &   &   & UH  & * & * &   &   \\ \hline
\cite{cheng2018big} &                                                  &   & * & * & * & * & * &   &   &   & UHR & * &   &   &   \\ \hline
\cite{cheng20175g} & *                                                & * & * & * &   & * & * & * & * &   & UHR & * & * &   & * \\ \hline
\cite{duan2016sdn} &                                                  & * & * & * &   & * & * &   & * &   & UH  &   & * &   & * \\ \hline
\cite{qi2018sdn} &                                                  & * & * & * &   & * &   &   &   &   & UH  & * & * &   &   \\ \hline
\cite{liu2017high} & *                                                & * & * & * & * & * & * & * & * & * & UH  & * & * &   & * \\ \hline
\cite{storck20195g} & *                                                & * & * & * & * & * & * & * & * & * & UH  & * & * &   & * \\ \hline
\cite{din20195g} &                                                  & * & * & * &   & * & * &   & * &   & UH  & * & * &   &   \\ \hline
\cite{aadil2018clustering} & *                                                & * & * & * &   & * &   &   & * &   & U   & * &   &   &   \\ \hline
\cite{duan2017sdn}n & *                                                & * & * & * &   & * & * &   &   &   & UH  & * & * &   &   \\ \hline
\cite{vladyko2019distributed} & *                                                & * & * & * &   & * & * &   &   &   & U   & * & * &   & * \\ \hline
\cite{alghamdi2020novel} & *                                                & * & * & * &   & * & * &   & * &   & U   & * &   &   &   \\ \hline
\cite{soua2018multi} & *                                                & * & * & * &   & * &   &   & * & * & UH  & * & * &   & * \\ \hline
\cite{ning2019vehicular} & *                                                & * & * & * &   & * & * & * & * &   & UH  & * &   &   & * \\ \hline
\cite{khattak2019integrating} &                                                  & * & * & * &   & * & * &   &   &   & UH  & * &   &   & * \\ \hline
\cite{bitam2015vanet} & *                                                & * & * &   &   & * & * &   & * & * & UH  & * &   &   & * \\ \hline
\end{tabular}%
}
\end{table*}


\begin{table*}[hbt!]
\begin{center}

\caption{Comparison between papers in 7 different metrics.
(N.M=Not Mentioned)}
\centering
\resizebox{\textwidth}{!}{%
\begin{tabular}{|c|c|c|c|c|c|c|c|}
\hline
\textbf{} &
  \textbf{\begin{tabular}[c]{@{}c@{}}Architecture type\\ Centralized/Distributed\end{tabular}} &
  \textbf{DSRC} &
  \textbf{Cellular} &
  \textbf{\begin{tabular}[c]{@{}c@{}}Incentive\\ mechanisim\end{tabular}} &
  \textbf{\begin{tabular}[c]{@{}c@{}}Data\\ Offloading\end{tabular}} &
  \textbf{V2V} &
  \textbf{V2I} \\ \hline
\textbf{\cite{han2018parallel}} & \textbf{C}    & \textbf{N.M} & \textbf{N.M} & \textbf{}  & \textbf{}  & \textbf{*} & \textbf{*} \\ \hline
\textbf{\cite{nitti2014adding}} & \textbf{C\&D} & \textbf{*}   & \textbf{*}   & \textbf{}  & \textbf{}  & \textbf{*} & \textbf{*} \\ \hline
\textbf{\cite{arnaboldi2014cameo}} & \textbf{C}    & \textbf{*}   & \textbf{*}   & \textbf{}  & \textbf{}  & \textbf{*} & \textbf{}  \\ \hline
\textbf{\cite{iamnitchi2012social}} & \textbf{C\&D} & \textbf{N.M} & \textbf{N.M} & \textbf{}  & \textbf{}  & \textbf{*} & \textbf{*} \\ \hline
\textbf{\cite{rahim2019social}} & \textbf{D}    & \textbf{N.M} & \textbf{N.M} & \textbf{}  & \textbf{}  & \textbf{*} & \textbf{}  \\ \hline
\textbf{\cite{paranjothi2016mavanet}} & \textbf{C}    & \textbf{*}   & \textbf{*}   & \textbf{}  & \textbf{}  & \textbf{*} & \textbf{*} \\ \hline
\textbf{\cite{mahdi2017grouping}} & \textbf{C\&D} & \textbf{N.M} & \textbf{N.M} & \textbf{}  & \textbf{}  & \textbf{*} & \textbf{*} \\ \hline
\textbf{\cite{gainaru2009realistic}} & \textbf{C}    & \textbf{N.M} & \textbf{N.M} & \textbf{}  & \textbf{}  & \textbf{}  & \textbf{*} \\ \hline
\textbf{\cite{liu2012exploring}} & \textbf{D}    & \textbf{*}   & \textbf{*}   & \textbf{}  & \textbf{}  & \textbf{*} & \textbf{}  \\ \hline
\textbf{\cite{huang2014social}} & \textbf{D}    & \textbf{*}   & \textbf{N.M} & \textbf{}  & \textbf{}  & \textbf{*} & \textbf{}  \\ \hline
\textbf{\cite{alam2014vedi}} & \textbf{C\&D} & \textbf{*}   & \textbf{*}   & \textbf{}  & \textbf{}  & \textbf{*} & \textbf{}  \\ \hline
\textbf{\cite{hu2012vssa}} & \textbf{C}    & \textbf{*}   & \textbf{*}   & \textbf{}  & \textbf{}  & \textbf{*} & \textbf{*} \\ \hline
\textbf{\cite{raza2018social}} & \textbf{C\&D} & \textbf{*}   & \textbf{*}   & \textbf{}  & \textbf{}  & \textbf{*} & \textbf{*} \\ \hline
\textbf{\cite{ning2017cooperative}} & \textbf{D}    & \textbf{*}   & \textbf{*}   & \textbf{}  & \textbf{}  & \textbf{*} & \textbf{}  \\ \hline
\textbf{\cite{zia2019towards}} & \textbf{D}    & \textbf{N.M} & \textbf{N.M} & \textbf{}  & \textbf{}  & \textbf{*} & \textbf{}  \\ \hline
\textbf{\cite{maglaras2016social}} & \textbf{D}    & \textbf{*}   & \textbf{*}   & \textbf{}  & \textbf{}  & \textbf{*} & \textbf{*} \\ \hline
\textbf{\cite{luo2019software}} & \textbf{C}    & \textbf{*}   & \textbf{*}   & \textbf{}  & \textbf{}  & \textbf{*} & \textbf{*} \\ \hline
\textbf{\cite{cheng2018big}} & \textbf{C}    & \textbf{*}   & \textbf{}    & \textbf{}  & \textbf{}  & \textbf{*} & \textbf{}  \\ \hline
\textbf{\cite{cheng20175g}} & \textbf{C}    & \textbf{*}   & \textbf{*}   & \textbf{}  & \textbf{}  & \textbf{*} & \textbf{*} \\ \hline
\textbf{\cite{duan2016sdn}} & \textbf{D}    & \textbf{*}   & \textbf{*}   & \textbf{}  & \textbf{}  & \textbf{*} & \textbf{*} \\ \hline
\textbf{\cite{qi2018sdn}} & \textbf{D}    & \textbf{*}   & \textbf{*}   & \textbf{}  & \textbf{}  & \textbf{*} & \textbf{*} \\ \hline
\textbf{\cite{liu2017high}} & \textbf{D}    & \textbf{*}   & \textbf{*}   & \textbf{}  & \textbf{}  & \textbf{*} & \textbf{*} \\ \hline
\textbf{\cite{storck20195g}} & \textbf{D}    & \textbf{}    & \textbf{*}   & \textbf{}  & \textbf{}  & \textbf{*} & \textbf{*} \\ \hline
\textbf{\cite{din20195g}} & \textbf{C}    & \textbf{}    & \textbf{*}   & \textbf{}  & \textbf{}  & \textbf{}  & \textbf{*} \\ \hline
\textbf{\cite{aadil2018clustering}} & \textbf{D}    & \textbf{*}   & \textbf{*}   & \textbf{*} & \textbf{}  & \textbf{*} & \textbf{*} \\ \hline
\textbf{\cite{duan2017sdn}} & \textbf{C}    & \textbf{*}   & \textbf{*}   & \textbf{}  & \textbf{}  & \textbf{*} & \textbf{*} \\ \hline
\textbf{\cite{vladyko2019distributed}} & \textbf{D}    & \textbf{*}   & \textbf{}    & \textbf{}  & \textbf{*} & \textbf{*} & \textbf{*} \\ \hline
\textbf{\cite{ge2016vehicular}} & \textbf{D}    & \textbf{N.M} & \textbf{N.M} & \textbf{}  & \textbf{*} & \textbf{}  & \textbf{*} \\ \hline
\textbf{\cite{alghamdi2020novel}} & \textbf{D}    & \textbf{*}   & \textbf{*}   & \textbf{}  & \textbf{}  & \textbf{*} & \textbf{*} \\ \hline
\textbf{\cite{khan2019two}} & \textbf{D}    & \textbf{*}   & \textbf{*}   & \textbf{}  & \textbf{}  & \textbf{*} & \textbf{}  \\ \hline
\textbf{\cite{song2014handover}} & \textbf{C}    & \textbf{}    & \textbf{*}   & \textbf{}  & \textbf{}  & \textbf{}  & \textbf{*} \\ \hline
\textbf{\cite{soua2018multi}} & \textbf{C\&D} & \textbf{N.M} & \textbf{N.M} & \textbf{}  & \textbf{}  & \textbf{*} & \textbf{*} \\ \hline
\textbf{\cite{ning2019vehicular}} & \textbf{C\&D} & \textbf{N.M} & \textbf{N.M} & \textbf{*} & \textbf{}  & \textbf{*} & \textbf{*} \\ \hline
\textbf{\cite{khattak2019integrating}} & \textbf{D}    & \textbf{*}   & \textbf{*}   & \textbf{}  & \textbf{*} & \textbf{*} & \textbf{*} \\ \hline
\textbf{\cite{bitam2015vanet}} & \textbf{C}    & \textbf{*}   & \textbf{*}   & \textbf{}  & \textbf{*} & \textbf{*} & \textbf{*} \\ \hline
\end{tabular}%
}
\end{center}
\end{table*}

\section{Research Challenges}
\label{challenges}

This section discusses the recent challenges in fog computing in VANET and provides future directions for dealing and addressing those challenges.

\paragraph{Need for SLA in fog architecture:}
Current Service Level Agreements or SLA have already been designed for cloud systems. As we have a heterogeneous environment in a fog system, there is a need to have a specific SLA that considers different requirements of fog systems and be compatible with them. 

\paragraph{Features aware fog system model:}
Many current system models that have been proposed for fog systems consider few objectives to be satisfied, such as the cost of the services, or QoS. It is important to take all the features in the fog system, whenever we want to propose a model. As a fog model, it's reasonable to consider many objectives such as: fast handover mechanism, sufficient bandwidth, quick response, low delay, and so on. Another important issue is that, in considering different features, usually bandwidth saving is being ignored that can have an important effect on the overall performance of the network.

\paragraph{Scalability issues:}
Another vital feature that most of the time has been neglected is scalability. There is a need to implement a system to be adoptable in business requirements, workload, system costs and so on. Because of the data explosion and the increasing number of sensors, this factor seems vital to be considered.

\paragraph{Mobile fog servers:}
As we discussed earlier, there is a definition of mobile fog computing. Mobile fog computing can be beneficial in many terms in the scope of VANET. Suppose that each car can act as a mobile fog node and integrating the number of driving cars can give us a powerful mobile fog server that can save cost, resources availability in the rural areas, improving Qos, and many other benefits. Although many papers consider mobile IoT devices, there are a few papers that discuss mobile fog computing servers. That’s why this term has remained in theory. While mobile fog servers can be an interesting and challenging concept.

\paragraph{System monitoring:}
Monitoring multiple or individual fog nodes are vital in the context of VANET. Suppose that in the network, if there would be a system to monitor and evaluate the performance of each fog node and compute the maximum number of connected devices to each fog node, we can have better QoS and QoE in our network, as we can be informed about each fog server status before happening congestion and interference. 

\paragraph{New standard in SDN for fog systems:}
SDN is one of the useful and beneficial techniques that are practical in large data centres. As it knows about dividing the control plane and data plane and efficiently manage both of them. Synchronization between many SDN controllers that are executed in fog servers seems challenging. So, having a new standard for SDN in fog models that support different domain types and different providers is essential.

\paragraph{High support for fast speed cars:}
Current technologies that have been proposed for fog systems in most cases cannot guarantee Qos for high-speed cars, as they miss their connection with different fog servers that have been implemented along the roads. Moreover, quick and proper handover management and offloading tasks for fast speed cars cannot be satisfied with them. Apart from that, the drivers and passengers that are in these cars cannot get desired services from fog computing servers as well. One of the interesting challenges for dealing with it is using historical data and machine learning schemes to predict their near future locations to switch between fog resources in advance.

\paragraph{Controlling peer to peer manner in fog:} Computing through incentive mechanisms
Unlike cloud nodes that can manage offloading tasks who are coming from fog nodes and different users, in fog servers, this task does not seem easy and straight forward, especially when two fog nodes decide to have offloading tasks between each other. This issue is challenging, as there are no other third parties in between to manage and control synchronization, available resources, handshaking mechanisms, and so on. Cooperation between fog nodes is vital when there are natural disasters, such as earthquakes or flooding. In this case, there is no cloud entity and the only nodes which can cooperate and manage the data are fog nodes. So, it is important to have some incentive mechanisms between mobile fog nodes for having better data dissemination. It is important to say that, in VSN, some selfish nodes can be categorized in two groups, individual and social. A node is selfish, when it only follows its interest and does not care about any other requests. A socially selfish node can be defined as a node who is looking for other nodes that have the same interests and only willing to cooperate with them.

\paragraph{Having a standard model for VSN:}
There are a few models that address VSN. Having an effective model that considers all the requirements for having a strong VSN like OSN seems crucial, if we want to have a stable and standard social network in the context of VANET. For example, how to deal with congestion, handover, and data offloading between different nodes in a social group of vehicles.

\paragraph{Assessment in a real-world with real data:}
Many applications are proposed in the context of VANET environment. However, most of them have been tested in  simulation environments, not the real world. Many of the applications that have been evaluated in the simulators will be confronted with many new challenges when we put them in the real world with real data. The scenario in the real world most of the time is not predictable. So, it's important to have real data for having better performance in our applications.

\section{Conclusion}
\label{conc}
 In this paper, we explained about fog and cloud computing and the main differences between them in the scope of VANET. After, we clarified the main advantages of Vehicular Social Network and how we can take benefits through using VSN in fog computing. Also, the main challenges in fog and cloud in VANET has been defined. Moreover, we fully explained how Vehicular Social Network as an emerging technology, can help the system for having stable connection. Also, by fully reviewing different novel papers in fog computing and VSN in VANET, we clarified research challenges and trends in using fog and VSN in VANET.

\printcredits

\bibliographystyle{cas-model2-names}

\bibliography{main.bib}


\bio{F2.jpg}
Farimasadat Miri received her B.S. and M.S. degree in information communication technology (ICT) in 
Iran University of Science and Technology. Currently, she is completing her Ph.D. degree at Ontario Tech University.
Her research field includes fog computing, Cloud computing, Vehicular ad hoc network and using AI in fog and cloud computing for resource allocation 
and optimization in smart city.
\endbio

\bio{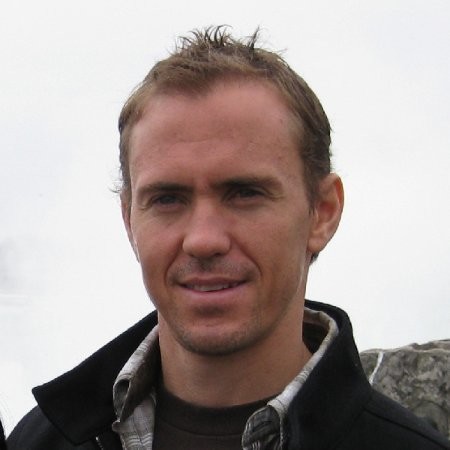}
Richard Pazzi is an Associate Professor at Ontario Tech University (former University of Ontario Institute of Technology), Canada. His research interests include fault-tolerant protocols for Wireless Sensor Networks and Mobile Computing. He is also active in the areas of Vehicular Ad Hoc Networks, multimedia communications and networked 3D virtual environments. He has served as Chair and TPC in a number of IEEE/ACM conferences. He is the recipient of Best Research Paper Awards from the IEEE Symposium on Computers and Communications (ISCC 2015), the Fifth International Conference on Advances in Vehicular Systems (VEHICULAR 2016), the IEEE International Conference on Communications (ICC 2009) and the International Wireless Communications and Mobile Computing Conference (IWCMC 2009), recipient of Elseviers Top Cited Article (2005−2010) for his work published in the Journal of Parallel and Distributed Computing (JPDC 2006). Contact him at richard.pazzi@uoit.ca
\endbio

\end{document}